\pdfoutput=1

\documentclass[twocolumn,aps,prapplied,superscriptaddress,longbibliography]{revtex4-2}
\usepackage{amssymb,amsfonts,amsmath,amsthm}
\usepackage{color,graphicx,hyperref,mathrsfs}
\usepackage{enumerate}
\usepackage{babel}
\usepackage[normalem]{ulem} \usepackage{siunitx,upgreek}
\usepackage{comment} 
\newcommand{\R}{\mathbb R} 
\newcommand{\q}{\mathbf q} \newcommand{\rr}{\mathbf r}

\newcommand{\m}{\mathbf{m}}
\newcommand{\mpa}{m^\parallel}
\newcommand{\Mpa}{M^\parallel}

\newcommand{\Mpe}{\mathbf{M}^\perp}

\newtheorem{theorem}{Theorem}

\begin{document}
\title{Optimal skyrmion stability in antisymmetric ultrathin
  ferromagnetic bilayers}

\author{Anne Bernand-Mantel}
\email{anne.bernand-mantel@cemes.fr}

\affiliation{Université de Toulouse, CNRS, CEMES,  Toulouse, France
}

\author{Valeriy V. Slastikov}

\affiliation{School of Mathematics, University of Bristol, Bristol BS8
  1UG, United Kingdom}

\author{Cyrill B. Muratov}

\affiliation{Dipartimento di Matematica, Universit\`a di Pisa, Largo
  B. Pontecorvo, 5, 56127 Pisa, Italy}

\date{\today}


\begin{abstract}
  We demonstrate the stray-field-mediated skyrmion stabilizing
  capabilities of ultrathin exchange-decoupled antisymmetric
  ferromagnetic bilayers based on conventional transition metal
  materials.  Using an asymptotically exact micromagnetic model valid
  in the ultrathin film limit, we show that the antisymmetric
  tailoring of the bilayer allows the Dzyaloshinskii-Moriya
  interaction and the dipolar interaction to act synergistically to
  stabilize skyrmions, in contrast to the monolayer case, in which
  these energies compete.  To obtain optimal stability of these
  skyrmions against collapse and bursting -- the two fundamental
  processes determining skyrmion lifetime, we carry out an asymptotic
  analysis of the saddle point solution that separates the skyrmion
  from the demagnetized state. The result is an optimal stability line
  for compact skyrmions in the non-dimensional parameter space of the
  effective Dzyaloshinskii-Moriya interaction strength and the
  effective film thickness. Our predictions are confirmed by extensive
  micromagnetic simulations of antisymmetric bilayers, using magnetic
  parameters of the conventional Pt/Co/AlO$_x$ systems. Our results
  provide a new pathway for experimental observations of 10 nm radius
  zero-field skyrmions with lifetimes compatible with information
  technology applications.
\end{abstract}

\maketitle

\section{Introduction}
\label{sec:introduction}

Magnetic skyrmions are topological solitons that are considered to be
promising candidates for ultra-dense and energy-efficient information
technology and unconventional computing applications
\cite{fert13,fert17,zhang20,lee23,mishra25}. The past decade has seen
major theoretical and experimental efforts from the research community
in material optimization towards observation of smaller and more
stable magnetic skyrmions. Despite these efforts, a reproducible
experimental realization of 10 nm isolated skyrmions at zero applied
magnetic field with lifetimes exceeding several seconds at room
temperature remains elusive.
  
In single ultrathin ferromagnetic layers with broken inversion
symmetry, compact skyrmions with diameters down to 1 nm have been
reported under magnetic fields of a few Tesla in PdFe/Ir(111) at
liquid helium temperature \cite{romming13}. More recently, a 5 nm
diameter skyrmion was observed at 4.2K in Rh/Co/Ir(111) at zero
applied magnetic field \cite{meyer19}. A way to enhance the skyrmion
lifetime in order to make them observable at room temperature (RT) is
to stack several asymmetrically capped ferromagnetic layers, as
observed in the Ir/Co/Pt system where skyrmions with diameters down to
30 nm were recorded at RT \cite{moreau-luchaire16}. In these
multilayer systems, the stray field plays a non-negligible role in the
stabilization of the observed skyrmionic bubbles
\cite{legrand18}. This prevents the reduction of the size below a
certain value due to the existence of an energy saddle that separates
the compact skyrmion, usually unstable at RT, from the experimentally
observed skyrmionic bubble \cite{buttner18,bernand-mantel18}. The
current-driven dynamics of these skyrmionic bubbles was also found to
be hindered by disorder \cite{woo16,legrand17}, as well as by the
non-uniformity of the magnetization in the direction perpendicular to
the sample plane \cite{legrand18}.

Another general issue related to the dynamics of skyrmions is their
tendency to move non-collinearly with the applied electric current
\cite{jiang17}, a phenomenon referred to as skyrmion Hall effect that
was predicted in the early works on skyrmions \cite{ivanov90}.  A
promising direction to reduce the skyrmion Hall effect is to use
rare-earth transition metal ferrimagnets, in which the gyroscopic
force generated by each magnetic sublattice is canceled at the
compensation temperature \cite{woo18,berges23}. Although high
current-induced skyrmion velocities have been reported in such films
\cite{caretta18,quessab22,berges23}, this system seems to generally
suffer from the same limitation as ferromagnetic thin films, namely,
the difficulty to stabilize compact skyrmions at room temperature and
zero applied magnetic field, and there exists only a single report of
sub 20 nm diameter compact skyrmions \cite{caretta18} that has not
been later reproduced.

The attention has recently shifted towards another system: the
synthetic antiferromagnet (SAF) system made of two ferromagnetic
layers with broken inversion symmetry, coupled antiferromagnetically
via a metallic non-magnetic spacer. This system offers the same
advantages as the compensated ferrimagnets in terms of high
current-induced skyrmion velocity and vanishing skyrmion Hall effect
\cite{dohi19,mallick24,pham24,jiang25,bhukta25}, as well as a
possibility to reach the flow regime for skyrmion current-induced
motion, as was recently demonstrated
\cite{mallick24,pham24,bhukta25}. On the other hand, issues have
emerged in this system such as the coexistence of ferromagnetic and
antiferromagnetic phases \cite{bran09,barker24,sim24} and the
existence of skyrmion inertia and a maximum velocity limit related to
the finite interlayer antiferromagnetic coupling \cite{panigrahy22}.
In addition, the property of SAF having a vanishing stray field due to
the antiparallel alignment of the magnetization vector in the two
layers makes the experimental observation of skyrmions even more
challenging than in the case of single ferromagnets and
ferromagnetically coupled multilayers. In a fully compensated SAF with
no stray field, skyrmionic bubbles are ruled out and only zero-field
compact skyrmions should be observed. In theory, stacking several
layers of compensated SAF could lead to an increase of the collapse
barrier of individual compact skyrmions \cite{legrand20,pham24}, while
not increasing their size. However, the experimentally observed
skyrmions remain in a few hundred nanometer to a few micron diameter
range and suffer from stripe-out instabilities
\cite{dohi19,mallick24,pham24,jiang25}.  This reveals their skyrmionic
bubble nature caused by the presence of residual stray fields
\cite{legrand20,finco25} due to the finite thickness of the layers and
the finite distance separating them. This issue has not been resolved,
despite significant efforts in material optimization
\cite{legrand20,chen24}.

\begin{figure}[t]
  \centering
\includegraphics[width=2.75in]{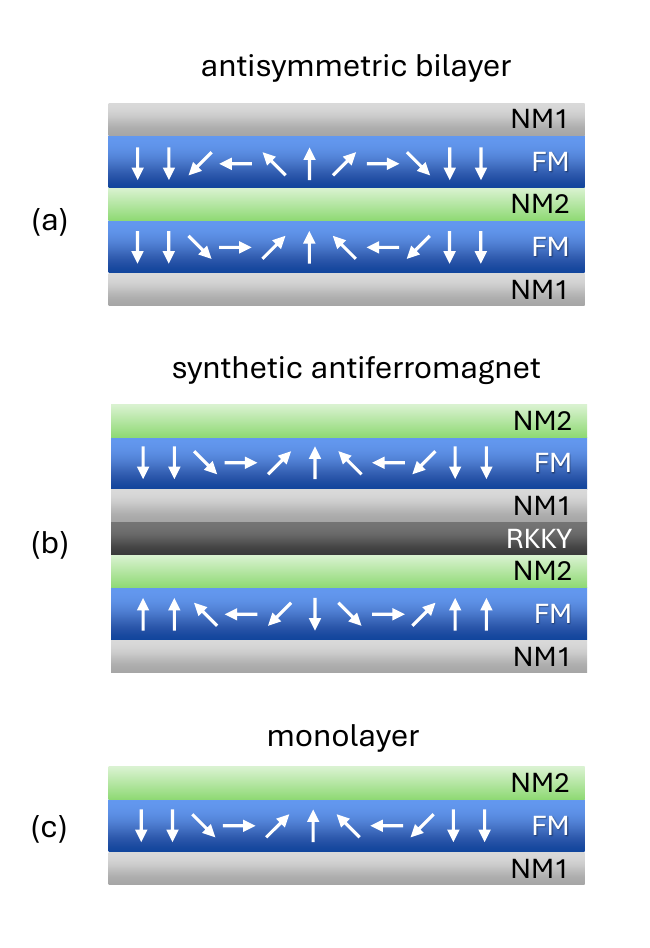}  
\caption{Schematics of the ferromagnetic multilayer structures with a
  skyrmion present: (a) antisymmetric ferromagnetic bilayer consisting
  of two identical ferromagnetic layers separated by a single
  non-magnetic layer (NM2) and capped by a different non-magnetic
  material (NM1); (b) a SAF consisting of two ferromagnetic layers
  (FM) capped by two different non-magnetic layers (NM1 and NM2) and
  separated by an exchange coupler (RKKY); (c) single ferromagnetic
  monolayer (FM) capped by two different non-magnetic layers (NM1 and
  NM2).}
  \label{fig:anti}
\end{figure}

The effect of stray field on domain walls and skyrmionic bubbles in
multilayers is a well studied phenomenon that gives rise to a
magnetization twist in the out-of-plane direction in multilayers
\cite{schloemann73,fallon18,lucassen19,legrand18,legrand18a,lemesh18}
and antiparallel domain wall configurations in bilayers
\cite{garcia06,bellec10}.  In a recent study, we demonstrated that in
ultrathin multilayers the interlayer dipolar interaction is taking the
form of a DMI-like energy term which contributes to this twist
\cite{dsbm:jns25} (see also \cite{lemesh18,dejong26}).  Our result was
applied to the case of stray field-coupled ferromagnetic bilayers in
the absence of DMI, for which we obtained minimizers in the form of a
pair of Néel skyrmions with opposite chiralities fixed by the
flux-closure pattern.  Exploiting this stray-field-minimizing
configuration to stabilize skyrmions in antisymmetric ferromagnetic
bilayers with opposite DMI remains largely unexplored. To date, there
have been only two experimental studies reporting on skyrmionic
bubbles in this system \cite{hrabec17,dejong26}. In the most recent
work \cite{dejong26}, the contribution from the DMI-like energy term
of dipolar origin to the wall energy has been estimated with the help
of relevant analytical models \cite{lemesh18} and micromagnetic
simulations.

In the present work, we explore the potential of antisymmetric
ferromagnetic bilayers for the realization of compact and stable
skyrmions compatible with memory applications.  To this aim, we derive
a reduced thin-film micromagnetic energy functional describing the
magnetization configurations in stacked ultrathin ferromagnetic layers
and identify just two dimensionless parameters -- the reduced DMI
strength and the reduced film thickness, governing the system's
behavior. We next focus on the antisymmetric ferromagnetic bilayer
system, for which we demonstrate existence of skyrmionic states (pairs
of N\'eel skyrmions, one per layer) as local energy minimizers and
identify the approximate boundary of the parameter region beyond which
these skyrmion states undergo bursting and strip-out. We then locate
an approximate optimal stability curve in the two-dimensional
parameter space, over which the bursting energy barrier is equal to
the collapse barrier, signifying the parameters at which the skyrmion
is most stable with respect to thermal noise.

Starting with this curve, we then carry out a two-dimensional
parameter sweep to numerically obtain a detailed phase diagram of
skyrmion solutions existence, along with their radii and the energy
barriers as functions of the reduced DMI strength and the reduced film
thickness.  We repeat this process for ferromagnetic monolayers and
then carry out a comparison between antisymmetric ferromagnetic
bilayers, SAF and ferromagnetic monolayers. Our results show that, for
typical magnetic parameters of transition metal multilayers,
antisymmetric bilayers exhibit a wide range of conditions under which
room-temperature-stable magnetic skyrmions with radii on the order of
10 nm can exist. This is due to the synergistic action of the
interfacial DMI and the dipolar interactions in antisymmetric
exchange-decoupled ferromagnetic bilayers. The stability properties of
these skyrmions compare favorably with those of the skyrmions in SAF
and monolayers.

The paper is organized as follows. In Sec. \ref{sec:Model}, we present
the basic micromagnetic model and then carry out its thin fim
reduction, followed by non-dimensionalization. In Sec. \ref{sec:abl},
we present a detailed analysis of the reduced thin film micromagnetic
energy for antisymmetric bilayers. In Sec. \ref{sec:comp}, we carry
out an analogous study for SAF and single ferromagnetic monolayers and
compare the results with those for antisymmetric bilayers. Finally, in
Sec. \ref{sec:concl} we draw conclusions.


\section{Model}
\label{sec:Model}

Our starting point is the micromagnetic energy of a stack of identical
ferromagnetic layers in the presence of intralayer exchange,
perpendicular magnetocrystalline anisotropy (both of bulk and
interfacial origin), interfacial DMI and the magnetostatic interaction
\cite{landau8,hubert,dsbm:jns25}. Consider $N$ stacked layers of a
single ferromagnetic material of thickness $d$ separated by
non-magnetic layers of thickness $(a-1)d$, where $a > 1$ is a
geometric factor corresponding to the ratio of the total thickness of the stack unit to that of the ferromagnet.  The ferromagnet
is characterized by saturation magnetization $M_s$ (in A/m) and
exchange stiffness $A$ (in J/m). For $n = 1, \ldots, N$, the $n$-th
ferromagnetic layer is assumed to occupy the region of space
\begin{align}
  \label{eq:Omn}
  \Omega_n
  & = \big\{ (x, y, z) \in \R^3 \ : \notag \\
  & (x, y) \in \R^2, \ z \in [a d (n - 1), a d (n - 1) + d] \big\}.  
\end{align}
Each layer $\Omega_n$ has bulk uniaxial perpendicular
magnetocrystalline anisotropy constant $K_u$ (in J/m$^3$), interfacial
anisotropy constants $K_n^{s,\pm}$ (in J/m$^2$) on the top and bottom
surfaces $\partial \Omega_n^\pm$ of $\Omega_n$, and interfacial DMI
strengths $D_n^{s,\pm}$ (in J/m) on $\partial \Omega_n^\pm$,
respectively.

The energy of such a system takes the following form, in the SI units
\cite{dsbm:jns25}:
\begin{align}
  \label{eq:E3dM}
  & \mathcal E(\mathbf M)
    = \sum_{n=1}^N \int_{\Omega_n} \left\{ {A \over M_s^2} |\nabla
    \mathbf M|^2 + {K_u \over M_s^2 } |\Mpe|^2 \right\} d^3
    r \notag 
  \\
  & - \frac{\mu_0}{2} \sum_{n=1}^N \int_{\Omega_n} \left(
    \mathbf H_d \cdot \mathbf M + M_s^2 \right) d^3 r \notag
  \\
  & + \sum_{n=1}^N \left( {K_n^{s,-} \over M_s^2} \int_{\partial \Omega_-} 
    | \Mpe|^2  d^2 r + {K_n^{s,+} \over M_s^2}
    \int_{\partial \Omega_+}  | \Mpe|^2  d^2 r \right)
    \notag \\ 
  & - \sum_{n=1}^N  {D_n^{s,-} \over M_s^2 } \int_{\partial \Omega_n^-} 
    \left( \Mpa \nabla \cdot \Mpe - \Mpe
    \cdot \nabla \Mpa \right) d^2 r \notag \\
  & + \sum_{n=1}^N  {D_n^{s,+} \over M_s^2} \int_{\partial \Omega_n^+}
    \left( \Mpa \nabla \cdot \Mpe - \Mpe
    \cdot \nabla \Mpa \right) d^2 r. 
\end{align}
Here $\mathbf M = (\Mpe, \Mpa)$ is the magnetization
vector, whose in-plane component is $\Mpe \in \R^2$ and the
out-of-plane component is $\Mpa \in \R$ (the superscripts are with
respect to the direction of the easy axis $\hat{\mathbf z}$), with the
convention that $|\mathbf M| = M_s$ in
$\Omega = \bigcup_{n=1}^N \Omega_n$ and $|\mathbf M| = 0$ outside
$\Omega$. The first line in the energy represents the intralayer
exchange plus the bulk perpendicular magnetocrystalline anisotropy,
the second line gives the contribution of the magnetostatic energy of
the demagnetizing field $\mathbf H_d$ vanishing for $z \to \pm \infty$
that solves the stationary Maxwell's equations
\begin{align}
  \label{Maxwell}
  \nabla \cdot (\mathbf H_d + \mathbf M) = 0, \qquad \nabla \times
  \mathbf H_d = 0,
\end{align}
from which we subtracted the contribution of the ferromagnetic state
to avoid divergent integrals; the third line gives the
magnetocrystalline anisotropy contributions of the bottom and top
surfaces of each layer, and the last two lines give the respective DMI
contributions of those surfaces. In writing the latter, we used the
sign convention that yields the same DMI constants,
$D_n^{s,+} = D_n^{s,-}$ for the interfaces between the ferromagnet
with the same non-magnetic material on both sides of the
ferromagnet. Notice that $\mathcal E(\mathbf M) = 0$ if
$\mathbf M = \pm M_s \hat{\mathbf z}$ in $\Omega$. Also notice that
the gradient in the DMI contributions acts only in-plane, as the
integrations there are carried out over the planes parallel to the
$xy$-plane. Finally, for simplicity of the presentation we omitted the
Zeeman energy term that can be easily added back to \eqref{eq:E3dM},
if needed \cite{bfbsm:prb23}. Other generalizations of the model, such
as addition of the interlayer exchange coupling \cite{dubicki:phd}, or
variable layer thicknesses and material parameters, are also
straightforward.

We next carry out a non-dimensionalization of the energy in
\eqref{eq:E3dM} by measuring lengths in the units of the exchange
length $\ell_{ex} = \sqrt{A / K_d}$, where
$K_d = \frac12 \mu_0 M_s^2$, introducing the normalized magnetization
vector $\m(\mathbf r) = \mathbf M(\ell_{ex} \mathbf r) / M_s$ for
$\mathbf r \in \R^3$, and defining the dimensionless energy in the
units of $Ad$, i.e., $E(\m) = \mathcal E(\mathbf M) / (Ad)$. Then
\begin{align}
  \label{eq:E3dm}
  & E(\m)
    = {1 \over \delta} \sum_{n=1}^N \int_{\ell_{ex}^{-1} \Omega_n}
    \left\{ |\nabla  
    \m|^2 + Q_u |\m^\perp|^2 \right\} d^3
    r \notag 
  \\
  & - \frac{1}{\delta} \sum_{n=1}^N \int_{\ell_{ex}^{-1} \Omega_n}
    \left(  \mathbf h_d \cdot \m + 1 \right) d^3 r \notag
  \\
  & + \sum_{n=1}^N Q_n^{s,-} \int_{\ell_{ex}^{-1} \partial
    \Omega_-}  | \m^\perp|^2  d^2 r \notag \\
  & + \sum_{n=1}^N Q_n^{s,+} 
    \int_{\ell_{ex}^{-1} \partial \Omega_+}  | \m^\perp|^2  d^2 r 
    \notag \\ 
  & -  \sum_{n=1}^N \kappa_n^- \int_{\ell_{ex}^{-1} \partial \Omega_n^-}
    \left( \mpa \nabla \cdot \m^\perp - \m^\perp
    \cdot \nabla \mpa \right) d^2 r \notag \\
  & +  \sum_{n=1}^N \kappa_n^+ \int_{\ell_{ex}^{-1} \partial
    \Omega_n^+} \left( \mpa \nabla \cdot \m^\perp - \m^\perp
    \cdot \nabla \mpa \right) d^2 r,
\end{align}
where, as before, $\m = (\m^\perp, \mpa)$, with $\m^\perp \in \R^2$
and $\mpa \in \R$ being the in-plane and out-of-plane components of
$\m$, and we introduced the dimensionless film thickness
$\delta = d / \ell_{ex}$ and the dimensionless quantities
\begin{align}
   \label{eq:dQ}
  Q_u = {K_u \over K_d}, \quad
  Q_n^{s,\pm} = {K_n^{s,\pm} \over d K_d}, \quad \kappa_n^\pm =
  {D_n^{s,\pm} \over  
  d \sqrt{A K_d}},
\end{align}
corresponding to the bulk anisotropy strength, surface anisotropy
strengths and the DMI strengths, respectively. The rescaled
demagnetizing field $\mathbf h_d$ satisfies
\begin{align}
  \label{eq:hdU}
  \mathbf h_d = -\nabla U, \qquad \Delta U = \nabla \cdot \mathbf m,  
\end{align}
in the sense of distributions in $\R^3$, with $\nabla U - \m$
vanishing at infinity \cite{dmrs:sima20}.

The above model presents a formidable challenge to analysis and may in
general only be treated through large-scale three-dimensional
numerical simulations, using a number of available software packages
(see, e.g.,
\cite{garcia07,schrefl07,kritsikis14,fu16,leliaert18,excalibur}). Nevertheless,
this model may be considerably simplified in the case of ultrathin
ferromagnetic layers that are relevant for most spintronic
applications, where each material layer extends to thicknesses of only
a few interatomic distances. It corresponds to an assumption that the
total multilayer thickness is  smaller than the
exchange length in the material,  $N a d \lesssim \ell_{ex}$. In
this case the micromagnetic energy may be expanded in the powers of
$\delta$, keeping only the leading order terms (for a detailed
derivation, see \cite{dsbm:jns25}). The micromagnetic energy then
obeys $E(\m) \simeq E_N(\m_1, \ldots, \m_N)$, where
$\m_n: \R^2 \to \mathbb S^2$ are the normalized magnetization vector
fields in the $n$-th ferromagnetic layer that depend only on two
spatial coordinates, and the reduced thin film energy $E_N$ reads
\begin{align}
  \label{eq:EN}
  E_N& (\m_1, \ldots, \m_N)
       = \notag \\
     & \sum_{n=1}^N \int_{\R^2} \left\{ |\nabla \m_n|^2 + (Q_n - 1)
       |\m_n^\perp|^2 \right\} d^2 r  \notag \\
     & - \sum_{n=1}^N \int_{\R^2} 2 \kappa_n \m_n^\perp \cdot \nabla
       m_n^\| d^2 r \notag \\
     & - \delta \sum_{n=1}^{N-1} \sum_{k=n+1}^N \int_{\R^2}
       \left( 
       \m^\perp_n \cdot \nabla m_k^\|  - 
       \m^\perp_k  \cdot \nabla m_n^\| \right) \, d^2 r \notag \\
     & - \delta \sum_{n=1}^{N} \sum_{k=1}^{N} \int_{\R^2}\int_{\R^2} 
       \frac{ \nabla m_n^{\|}(\mathbf r) \cdot \nabla 
       m_k^{\|}(\mathbf r')
       }{4\pi|\mathbf r-\mathbf r'|}  d^2r\ d^2r' \notag \\
     & + \delta \sum_{n=1}^{N} \sum_{k=1}^{N} \int_{\R^2}\int_{\R^2}       
       \frac{ \nabla \cdot \m_n^\perp(\mathbf r) \nabla \cdot
       \m_k^\perp(\mathbf r') }{4\pi|\mathbf r-\mathbf r'|}  d^2r\ d^2r' ,
\end{align}
where in the second and fourth lines of the right-hand side we carried
out an integration by parts to rewrite the respective integrals (see
\cite{dms:mmmas24}). Here
\begin{align}
  \label{eq:Qkappa}
  Q_n = Q_u + Q_n^{s,+} + Q_n^{s,-}, \qquad \kappa_n =
  \kappa_n^+ - \kappa_n^-,
\end{align}
and the first line in the right-hand side of \eqref{eq:EN} contains
the contributions of the intralayer exchange and effective
perpendicular magnetic anisotropy that includes the local contribution
of the intralayer demagnetizing field (shape anisotropy), the second
line represents the DMI energy, and the rest of the terms are the
additional $O(\delta)$ contributions to the stray field energy. In
particular, the third line contains a stray field-mediated DMI-like
term from the interplay between the volume and the surface charges in
different layers that was identified in \cite{dsbm:jns25}. The fourth
and the fifth lines in the right-hand side of \eqref{eq:EN} represent,
respectively, the nonlocal contributions of the surface-surface and
volume-volume charge interactions to the stray field energy.

\section{Theory of antisymmetric bilayers}
\label{sec:abl}

We now turn our attention to the particular case of antisymmetric
bilayers. For this system, we have $N = 2$, $Q_1 = Q_2 = Q > 1$ and
$\kappa_1 = -\kappa_2 = \kappa \geq 0$, favoring a counter-clockwise
rotation in the bottom layer and clockwise rotation in the top layer,
with a flux closure-like arrangement of the magnetization in the two
layers \cite{dsbm:jns25}, see Fig. \ref{fig:anti}(a).

\subsection{Thin film energy}

The corresponding thin film energy is
\begin{align}
  \label{eq:E2}
  E_2
  & (\m_1, \m_2) = \sum_{n=1}^2 \int_{\R^2} \left( |\nabla \m_n|^2 +  
    (Q - 1) |\m_n ^\perp|^2 
    \right) d^2 r \notag \\
  & - 2 \kappa \int_{\R^2} \left( \m_1^\perp \cdot \nabla m_1^\| -
    \m_2^\perp \cdot \nabla m_2^\| \right) d^2 r \notag \\
  & - \delta \int_{\R^2} \left( \m_1^\perp \cdot \nabla m_2^\| -
    \m_2^\perp \cdot \nabla m_1^\| \right) d^2 r \notag \\
  & - \delta \sum_{n=1}^2 \sum_{k=1}^2 \int_{\R^2} \int_{\R^2}
    {\nabla m_n^\|(\mathbf r) \cdot \nabla m_k^\|(\mathbf r') \over 4
    \pi |\mathbf r - \mathbf r'|}  \, d^2r  \, d^2 r' \notag \\
  & + \delta \sum_{n=1}^2 \sum_{k=1}^2 \int_{\R^2} \int_{\R^2}
    {\nabla \cdot \m_n^\perp(\mathbf r) \nabla \cdot
    \m_k^\perp(\mathbf r') 
    \over 4 \pi |\mathbf r - \mathbf r'|}  \, d^2r \,  d^2 r'. 
\end{align}
Notice that this energy is invariant with respect to the scaling
transformation $\mathbf r \to \lambda \mathbf r$, provided that
\begin{align}
  \label{eq:scale}
  Q - 1 \to {Q - 1 \over \lambda^2}, \quad \kappa \to {\kappa \over
  \lambda}, \quad \delta \to {\delta \over \lambda},
\end{align}
which represents an important underlying scaling symmetry of the thin
film energy. This allows us to eliminate one of the dimensionless
parameters by introducing the new parameters
\begin{align}
  \label{eq:dkbar}
  \bar\delta = {\delta \over \sqrt{Q - 1}}, \qquad \bar\kappa =
  {\kappa \over \sqrt{Q - 1}},
\end{align}
leading to the following simplified dimensionless energy
\begin{align}
  \label{eq:E2b}
  \bar E_2
  & (\m_1, \m_2) = \sum_{n=1}^2 \int_{\R^2} \left( |\nabla \m_n|^2 +  
    |\m_n ^\perp|^2 
    \right) d^2 r \notag \\
  & - 2 \bar \kappa \int_{\R^2} \left( \m_1^\perp \cdot \nabla m_1^\| -
    \m_2^\perp \cdot \nabla m_2^\| \right) d^2 r \notag \\
  & - \bar \delta \int_{\R^2} \left( \m_1^\perp \cdot \nabla m_2^\| -
    \m_2^\perp \cdot \nabla m_1^\| \right) d^2 r \notag \\
  & - \bar \delta \sum_{n=1}^2 \sum_{k=1}^2 \int_{\R^2} \int_{\R^2}
    {\nabla m_n^\|(\mathbf r) \cdot \nabla m_k^\|(\mathbf r') \over 4
    \pi |\mathbf r - \mathbf r'|}  \, d^2r  \, d^2 r' \notag \\
  & + \bar \delta \sum_{n=1}^2 \sum_{k=1}^2 \int_{\R^2} \int_{\R^2}
    {\nabla \cdot \m_n^\perp(\mathbf r) \nabla \cdot
    \m_k^\perp(\mathbf r') \over 4 
    \pi |\mathbf r - \mathbf r'|}  \, d^2r \,  d^2 r',
\end{align}
that satisfies $E_2(\m_1, \m_2) = \bar E_2(\bar \m_1, \bar \m_2)$,
where $\bar \m_n(\mathbf r) = \m_n(\mathbf r / \sqrt{Q - 1})$. In
other words, $\bar E_2$ gives the energy of the normalized
magnetization configuration $(\m_1, \m_2)$ in the two layers, with
lengths measured in the units of the Bloch wall thickness
$L_B = \ell_{ex} / \sqrt{Q - 1}$. The behavior of the energy
$\bar E_2$ is completely determined by only two dimensionless
parameters: $\bar\kappa$ and $\bar\delta$.

\subsection{Antisymmetric profiles}
\label{sec:antisymm-prof}

To better understand the behavior of the critical points of the energy
in \eqref{eq:E2b}, we consider an ansatz in which the in-plane
components of the magnetization in the two layers are anti-parallel,
while the out-of-plane components are parallel. Setting
\begin{align}
  \label{eq:m12ansatz}
  \m_1 = (\m^\perp, \mpa), \qquad  \m_2 = (-\m^\perp, \mpa),
\end{align}
for $\m = (\m^\perp, \mpa)$ leads to
$\bar E_2(\m_1, \m_2) = 2 \bar E^\pm(\m)$, where
\begin{align}
  \label{eq:E2b0}
  \bar E^\pm
  & (\m) 
    = \notag \\
  & \int_{\R^2} \left( |\nabla \m|^2 + |\m^\perp|^2 - (2
    \bar \kappa + \bar\delta) \m^\perp \cdot \nabla \mpa \right) d^2 r
    \notag \\
  & - 2 \bar \delta \int_{\R^2} \int_{\R^2}
    {\nabla \mpa(\mathbf r) \cdot \nabla \mpa(\mathbf r') \over 4
    \pi |\mathbf r - \mathbf r'|}  \, d^2r  \, d^2 r'.
\end{align}
This ansatz is motivated by the energetic advantages it provides: in a
skyrmion profile the DMI and the surface-volume charge interactions of
both layers cooperate, while the energy penalty due to the
volume-volume charge interactions cancels out. In particular, the
ansatz has been rigorously shown to be valid in the conformal limit
for $\bar\kappa = 0$ \cite{dsbm:jns25} and is confirmed by the
numerical simulations for the parameter ranges at which the skyrmion
solutions are observed (see Sec. \ref{sec:numer-simul-antisymm}).

\subsection{Bilayer skyrmions}
\label{sec:bilay-skyrm}

The expression in \eqref{eq:E2b0} coincides with the energy of a
single ferromagnetic layer of thickness $2 \bar\delta$ with the
effective DMI constant $\bar \kappa + \frac12 \bar \delta$, in which
the volume-volume charge interaction penalty term is absent. One can
easily adapt the proof of existence of skyrmion solutions as local
energy minimizers with topological degree $q(\m) = 1$, where
\begin{align}
  \label{eq:q}
  q(\m) = {1 \over 4 \pi} \int_{\R^2} \m \cdot (\partial_x \m \times
  \partial_y \m) \, d^2 r
\end{align}
defines the Brouwer degree of a map $\m: \R^2 \to \mathbb S^2$, with
the convention that $\m(\infty) = -\hat{\mathbf z}$
\cite{bms:prb20,bms:arma21}.  More precisely, we have the following
existence result.

\begin{theorem}
  \label{t:1}
  Let $\bar\delta \geq 0$ and $\bar\kappa \geq 0$ be such that
  \begin{align}
    \label{condQd}
    0 <  2 \bar \kappa + 3 \bar\delta \leq \sqrt{2}.
  \end{align}
  Then there exists a minimizer of $\bar E^\pm$ among all
  $\mathbf m \in H^1_\mathrm{loc}(\R^2; \mathbb S^2)$ such that
  $q(\mathbf m) = 1$,
  $\int_{\R^2} |\nabla \mathbf m|^2 d^2 r < 16 \pi$, and
  $\mathbf m +\hat{\mathbf{z}} \in L^2(\R^2; \R^3)$.
\end{theorem}
\noindent As was already mentioned, the asymptotic behavior of these
solutions in the absence of the DMI, $\bar \kappa = 0$, and in the
conformal limit $\bar\delta \to 0$ has been explicitly identified in
\cite[Theorem 2]{dsbm:jns25}. This result may be straightforwardly
extended to the case of $\bar\kappa > 0$ tending jointly to zero,
yielding the following expression for the dimensionless radius
$\bar\rho_2^\mathrm{sky}$ of the skyrmion in an antisymmetric bilayer:
\begin{align}
  \label{eq:rhob2}
  \bar\rho_2^\mathrm{sky}  \simeq {32 \bar\kappa + (16 + \pi^2)
  \bar\delta \over -64 W_{-1}\left( -\frac{32 \bar\kappa + (16 +
  \pi^2)  \bar\delta }{128} e^{1 + \gamma} \right)},    
\end{align}
where $\gamma \approx 0.5772$ is the Euler-Mascheroni constant and
$W_{-1}(t)$ is the Lambert $W$ function \cite{corless96}, while the
skyrmion energy satisfies
\begin{align}
  \label{eq:Esky2}
  \bar E_2^\mathrm{sky}
  & \simeq 16 \pi \notag \\
  & - {\pi \bar\rho_2^\mathrm{sky} \over 4} \left( 32 \bar \kappa
    + (16 + \pi^2) \bar
    \delta - 32 \bar\rho_2^\mathrm{sky} \right).
\end{align}
The above two formulas are valid for
$0 < \bar\kappa + \bar \delta \ll 1$. The solutions are asymptotically
radial, satisfy \eqref{eq:m12ansatz} as minimizers of \eqref{eq:E2b}
(as follows from the argument in \cite{dsbm:jns25}), and are
conjectured to be such for all values of the parameters.

\subsection{Skyrmion bursting}
\label{sec:skyrmion-bursting}

We now investigate the range of existence of the above solutions as
the values of $\bar\kappa$ and $\bar\delta$ are increased beyond the
range covered by Theorem \ref{t:1}. Assuming radial symmetry, for
$\bar\delta = 0$ such an analysis was performed in \cite{komineas21},
where it was shown, using formal asymptotic analysis and numerical
simulations, that as $\bar\kappa$ approaches the critical value
$\bar\kappa_c = 4 / \pi$ from below, the skyrmion solution transforms
into a bubble-like profile $\m =\m_{\bar\rho}$, where
\begin{align}
  \label{eq:bubble}
  \m_{\bar\rho}(\rr) \simeq  \left( -{\rr \over |\rr|} \mathrm{sech}
  (|\rr| - {\bar\rho}),  \tanh ({\bar\rho} - |\rr|) \right), 
\end{align}
whose radius ${\bar\rho}$ is determined by the value of $\bar\kappa$
and diverges as $\bar\kappa \to \bar\kappa_c$. The analysis of
\cite{komineas21} yields
$\bar \rho \simeq \sqrt{c \bar \kappa_c \over 2 (\bar \kappa_c - \bar
  \kappa)}$, with the constant $c \approx 0.9605$ determined
numerically. Nevertheless, as was already pointed out in
\cite{komineas21}, this radius may be obtained with a very good
accuracy (yielding $c = 1$ above) by plugging in the ansatz in
\eqref{eq:bubble} into the energy and calculating its local maximum in
${\bar\rho}$ to the leading order in $\bar \rho \gg 1$, as was done in
\cite{rohart13}. Below we use the same approximation in the presence
of the non-local terms in the energy to investigate the behavior of
solutions for $\bar\kappa \sim \bar\kappa_c$ and $\bar\delta > 0$.

\begin{figure}[t]
  \centering
  \includegraphics[width=3.25in]{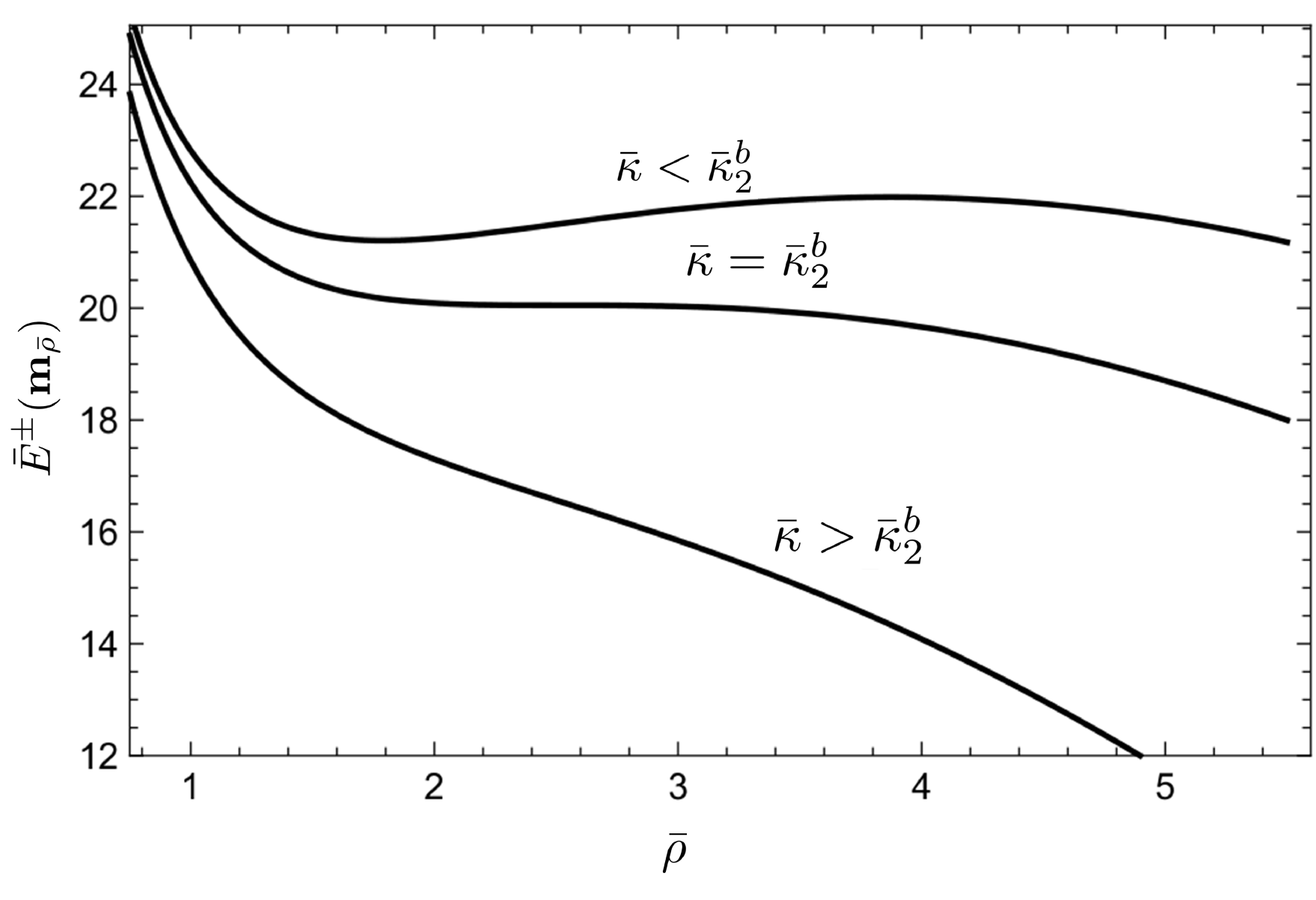}
  \caption{The dependence of $\bar E^\pm(\m_{\bar \rho})$ from
    \eqref{eq:EpmR} on $\bar \rho$ for $\bar\delta = 0.5$ and
    $\bar \kappa = 0.4$, $\bar \kappa = 0.4293$, and
    $\bar \kappa = 0.5$, respectively.}
  \label{fig:Epmrho}
\end{figure}

Substituting the ansatz in \eqref{eq:bubble} into the energy in
\eqref{eq:E2b0} and calculating the stray field energy in Appendix
\ref{sec:appstray}, we obtain
\begin{align}
  \label{eq:EpmR}
  \bar E^\pm(\m_{\bar\rho})
  & \simeq 2 \pi {\bar\rho} \left[ 4 - \pi (\bar\kappa + \tfrac12
    \bar\delta) \right] + {4 \pi \over {\bar\rho}} \notag \\
  & \quad - 8 \bar \delta  {\bar\rho} \ln
    \left(\frac{8 {\bar\rho}}{\pi } e^{\gamma - 1} \right),
\end{align}
to the leading order in ${\bar\rho} \gg 1$, with relative errors of
algebraic order in $1/\bar \rho$. In this expression, the first term
is the domain wall energy for the bubble of radius ${\bar\rho}$, the
second term is an additional exchange energy contribution from the
angular gradient of the magnetization \cite{rohart13}, and the last
term is the contribution of surface-surface charge interaction. This
expression may be analyzed as a function of ${\bar\rho}$, yielding two
critical points for $\bar \kappa < \bar \kappa_2^b(\bar \delta)$, one
critical point for $\bar \kappa = \bar \kappa_2^b(\bar \delta)$, or no
critical points, for $\bar \kappa > \bar \kappa_2^b(\bar \delta)$, for
any fixed $\bar\delta > 0$ (see Fig. \ref{fig:Epmrho}). The borderline
case of one critical point occurs at the inflection point of the
energy $\bar E^\pm(\m_{\bar\rho})$ as a function of ${\bar\rho}$ given
by
${\bar\rho} = {\bar\rho}_2^b = \left( \pi / \bar\delta
\right)^{1/2}$. A simple calculation then shows that
\begin{align}
  \label{eq:kappab2}
  \bar\kappa_2^b \simeq {4 \over \pi} - {2 \bar\delta \over \pi^2} \left[
  \ln
  \left( \frac {64}{\pi \bar
  \delta} \right) +1+2 \gamma + \frac{\pi^2}{4}\right].
\end{align}
This formula is expected to be accurate for $\bar\delta \lesssim 1$,
corresponding to our assumption ${\bar\rho} \gtrsim 1$, and gives the
predicted transition between the parameter region where
$\bar\kappa < \bar\kappa_2^b(\bar\delta)$, for which a local minimum
of the energy $\bar E^\pm(\m_{\bar\rho})$ exists, and a region where
$\bar\kappa > \bar\kappa_2^b(\bar\delta)$, for which no local minimum
exists and the energy is monotonically decreasing in
${\bar\rho}$. Thus, the value of $\bar\kappa_2^b$ may be associated
with a bursting line for the skyrmion solution in antisymmetric
ferromagnetic bilayers, with the bursting radius ${\bar\rho}_2^b$.

\subsection{Thermal stability}
\label{sec:thermal-stability}

We now turn to the characterization of thermal stability of the
skyrmion solutions for $\bar\delta > 0$ and
$0 < \bar\kappa < \bar\kappa_2^b(\bar\delta)$. Note that due to its
topological protection, a skyrmion may not continuously transform into
the uniformly magnetized ferromagnetic state. Nevertheless, as we
argued in \cite{bms:pnas22}, a skyrmion may disappear discontinuously
as a result of a singularity formation (at the continuum level)
leading to skyrmion collapse (see also
\cite{rohart16,buttner18,bernand-mantel18,heil19,bms:prb20}). The
energy cost of such a barrier-crossing event is determined by the
transition state in the form of a ``zero-radius'' skyrmion, which is
equal to the minimal value of the exchange energy in the considered
topological class. In the case of the antisymmetric bilayers with the
ansatz in \eqref{eq:m12ansatz} this results in the collapse barrier
\begin{align}
  \label{eq:dE2pmc}
  \Delta \bar E_2^{c,0} = 16 \pi - 2 \bar E^\pm(\m^\mathrm{sky}),
\end{align}
where $\m^\mathrm{sky}$ is the profile that locally minimizes
$\bar E^\pm$ among all $\m : \R^2 \to \mathbb S^2$ with $q(\m) = 1$
and $\m(\infty) = -\hat{\mathbf z}$.

At the same time, in the presence of the non-local effects due to
stray field another scenario is possible, in which the skyrmion radius
grows under the action of thermal noise, reaching the saddle point
solution corresponding to the largest critical point of the energy in
\eqref{eq:EpmR} for ${\bar\rho} > {\bar\rho}_2^b$. When $\bar\kappa$
is not in the immediate vicinity of $\bar\kappa_c$ and
$\bar\delta > 0$ is sufficiently small, we can drop the
$4 \pi / \bar\rho$ term from the energy in \eqref{eq:EpmR} and
calculate that the saddle point solution corresponds to the radius
$\bar\rho = \bar \rho_2^\mathrm{sad}$, where
\begin{align}
  \label{eq:R2s}
  \bar\rho_2^\mathrm{sad} \simeq {\pi \over 8} \exp \left(  {\pi \over
  \bar \delta} - {\pi^2 \bar\kappa \over 4 \bar\delta}  -
  \frac{\pi^2}{8} - \gamma \right).  
\end{align}
The associated energy barrier is
\begin{align}
  \label{eq:dEpms}
  \Delta \bar E_2^{b,0} = 2 \bar
  E^\pm(\m_{\bar\rho_2^\mathrm{sad}}) - 2 \bar 
  E^\pm(\m^\mathrm{sky}),
\end{align}
where
\begin{align}
  \label{eq:E2sad}
  \bar E^\pm(\m_{\bar \rho_2^\mathrm{sad}}) \simeq \pi \bar \delta
  \exp \left( {\pi \over \bar \delta} - {\pi^2 \bar \kappa \over 4 \bar
  \delta} - \gamma - {\pi^2 \over 8} \right).
\end{align}
These formulas are expected to be asymptotically exact for
$0 \leq \bar \kappa < \bar \kappa_c$ as $\bar \delta \to 0$.

Comparing the two energy barriers, one can see that the most likely
thermally activated event (under the ansatz in \eqref{eq:m12ansatz}
and the radiality assumption) would be determined by the minimum value
of the two barriers above, and the skyrmion lifetime in the Arrhenius
regime will, therefore, be given by
\begin{align}
  \label{eq:Arrh}
  \tau = \tau_0 \exp \left( {A d \over k_BT} \, \Delta \bar E_2 \right), 
\end{align}
where
$\Delta \bar E_2 = \min ( \Delta \bar E_2^{c,0}, \Delta \bar
E_2^{b,0})$, $k_B$ is the Boltzmann constant, $T$ is temperature, and
$\tau_0$ is the characteristic attempt timescale. It is thus possible
to maximize the lifetime by choosing the parameters of the problem for
which the two barriers above coincide, which amounts to setting
\begin{align}
  \label{eq:dEequal}
  \bar E^\pm(\m_{\bar\rho_2^\mathrm{sad}}) = 8 \pi.
\end{align}
This yields the predicted optimal value 
$\bar\kappa = \bar\kappa_2^\mathrm{opt}$ of the dimensionless DMI
strength
\begin{align}
  \label{eq:kappa2opt}
  \bar \kappa_2^\mathrm{opt} \simeq {4 \over \pi} - {4 \bar \delta
  \over \pi^2} \left[ \ln \left( {8  \over \bar \delta} \right) +
  \gamma + {\pi^2 \over 8} \right],
\end{align}
for fixed $\bar \delta > 0$.  Once again, this formula is expected to
be accurate for $\bar\delta \lesssim 1$, yielding
$\bar\rho_2^\mathrm{sad} \gtrsim 1$.

We wish to point out, however, that in reality the determination of
the barrier-crossing events in the case of bilayers is more
challenging due to the fact that the assumption of antisymmetry in
\eqref{eq:m12ansatz} used to derive the above simple expressions may
not be the most favorable for the transition states. More
specifically, the collapse barrier should not require that the
skyrmions in both layers collapse simultaneously, as the more
energetically favorable scenario should be the collapse in only one
layer. Thus, the value of $16 \pi$ in \eqref{eq:dE2pmc} should be
replaced with
$8 \pi + \bar E_2(\m_1^\mathrm{sky1}, \m_2^\mathrm{sky1})$, where
$8\pi$ is the energy of the zero-radius skyrmion in one layer and
$(\m_1^\mathrm{sky1}, \m_2^\mathrm{sky1})$ is the locally energy
minimizing configuration for $\bar E_2(\m_1, \m_2)$ with $q(\m_1) = 1$
and $q(\m_2) = 0$, i.e., the configuration with a skyrmion in only one
layer. Here the fact that the topologically non-trivial profile is
assumed to be in the first layer is not essential, since the energy
$\bar E_2(\m_1, \m_2)$ enjoys a symmetry with respect to swapping the
two layers, provided that the sign of the in-plane components is
reversed, i.e., we have
$\bar E_2(\m_1, \m_2) = \bar E_2(\tilde \m_1, \tilde \m_2)$, where
$\tilde \m_{1,2} = (-\m_{2,1}^\perp, m_{2,1}^\|)$. Hence, the collapse
barrier should rather be
\begin{align}
  \label{eq:dE2c}
  \Delta \bar E_2^c = 8 \pi + \bar E_2(\m_1^\mathrm{sky1},
  \m_2^\mathrm{sky1}) - \bar E_2 (\m_1^\mathrm{sky2},
  \m_2^\mathrm{sky2}), 
\end{align}
where $(\m_1^\mathrm{sky2}, \m_2^\mathrm{sky2})$ is the locally energy
minimizing configuration for $\bar E_2(\m_1, \m_2)$ with
$q(\m_1) = q(\m_2) = 1$, i.e., the configuration with a skyrmion in
both layers. Analogously, the bursting barrier is a priori given by
\begin{align}
  \label{eq:dEb2}
  \Delta \bar E_2^b = \bar E_2(\m_1^\mathrm{sad2}, \m_2^\mathrm{sad2}) -
  \bar E_2 (\m_1^\mathrm{sky2}, 
  \m_2^\mathrm{sky2}),  
\end{align}
where $(\m_1^\mathrm{sad2}, \m_2^\mathrm{sad2})$ is the lowest energy
saddle point configuration with
$q(\m_1^\mathrm{sad2}) = q(\m_2^\mathrm{sad2}) = 1$ connecting the
skyrmion configuration $(\m_1^\mathrm{sky2}, \m_2^\mathrm{sky2})$ with
the striped-out state via gradient descent.

In practice, the barriers with or without the antisymmetry ansatz in
\eqref{eq:m12ansatz} are expected to be close to one another, as both
the locally energy minimizing configuration
$(\m_1^\mathrm{sky2}, \m_2^\mathrm{sky2})$ with skyrmions in both
layers and the lowest energy saddle point configuration
$(\m_1^\mathrm{sad2}, \m_2^\mathrm{sad2})$ are expected to obey
\eqref{eq:m12ansatz}, while the energy of
$(\m_1^\mathrm{sky1}, \m_2^\mathrm{sky1})$ for the parameters of
optimal stability of $(\m_1^\mathrm{sky2}, \m_2^\mathrm{sky2})$ should
be close to that of the zero-radius skyrmion, i.e., one expects
$\bar E_2(\m_1^\mathrm{sky1}, \m_2^\mathrm{sky1}) \simeq 8 \pi$ for
most of the relevant parameters (see also
Sec. \ref{sec:numer-simul-antisymm}).

\subsection{Beyond bursting}
\label{sec:stripes}

As bursting of a skyrmion is expected to lead to the formation and
growth of a skyrmionic bubble followed by stripe-out, it is also
instructive to calculate the period of the resulting ground state
periodic stripe configuration corresponding to the demagnetizing state
in the considered system. This can be done exactly as in
\cite{bsm:prb25}, and with a few straightforward modifications we
obtain that the energy density per unit area (in the units of
$Ad/L_B^2$) of a periodic stripe configuration
$\m^\mathrm{stripe}_{\bar L}$ with period $\bar L$ (in the units of
$L_B$) in one layer is given by
\begin{align}
  \label{eq:Estripe}
  & {2 \bar E^\pm(\m^\mathrm{stripe}_{\bar L}) \over \bar L^2}
    \simeq {4 \over
    \bar L} \left[ 4 - \pi (\bar \kappa + \tfrac12 \bar \delta ) \right]
    \notag \\
  & \qquad \quad - {16 \bar \delta \over \pi \bar L} \left[ \ln \left(
    {\bar L \over 
    \pi^2} \right) + \gamma + 1 \right],
\end{align}
where we assumed that the configuration in the two layers satisfies
\eqref{eq:m12ansatz} with $\m = \m^\mathrm{stripe}_{\bar
  L}$. Minimizing this expression with respect to $\bar L$ yields the
optimal stripe period
\begin{align}
  \label{eq:Lopt}
  \bar L_2^\mathrm{opt} \simeq \pi^2 \exp \left( \dfrac{\pi}{\bar\delta} -
  \dfrac{\pi^2 \bar\kappa}{4 \bar\delta} - \dfrac{\pi^2}{8} - \gamma
  \right) . 
\end{align}
Again, these formulas are expected to be asymptotically exact for
$0 \leq \bar \kappa < \bar \kappa_c$ as $\bar \delta \to 0$.

Comparing the diameter of the saddle point solution obtained in
Sec. \ref{sec:thermal-stability} and the equilibrium width of a single
stripe, we see that their ratio reaches a universal value:
\begin{align}
  \label{eq:rsLopt}
  {2 \bar \rho_2^\mathrm{sad} \over \frac12 \bar L_2^\mathrm{opt}}
  \simeq {1 \over 2 \pi} \approx 0.1592
\end{align}
for $\bar\delta \ll 1$ and $\bar \kappa$ not too close to
$\bar \kappa_c$.  Thus, there should be a fairly well-defined range of
radial growth of the skyrmionic bubble after bursting caused by
thermal fluctuations before it undergoes a stripe-out
instability. This is indeed observed numerically in sufficiently large
computational domains upon skyrmion bursting.

\subsection{Numerical skyrmion solutions and energy barriers}
\label{sec:numer-simul-antisymm}

We now perform a numerical construction of skyrmion solutions as local
energy minimizers in the $(\bar \kappa, \bar \delta)$ plane by
carrying out micromagnetic simulations of the antisymmetric bilayer
system, using the {\sc Mumax3} software \cite{leliaert18}. For that
purpose, we consider two adjacent exchange-decoupled ferromagnetic
layers of thickness $d = 1$ nm and fix the parameters of the magnetic
material to have the characteristic values for transition metal
ultrathin ferromagnetic layers such as Pt/Co/AlO$_x$ with $A = 20$
pJ/m and $M_s = 1$ MA/m \cite{belmeguenai15,eyrich14}. This yields the
exchange length $\ell_{ex} \approx 5.64$ nm. We next vary the DMI
strength $D = D^{s,+}/d$ (in the units of J/m$^2$, conventionally
normalized per unit volume), setting the DMI constant in the bottom
layer to $D$ and the DMI constant in the top layer to $-D$,
respectively, as well as the anisotropy constant $K_u$, in such a way
as to express the results in terms of the dimensionless quantities
$\bar\kappa$ and $\bar\delta$ defined in \eqref{eq:dkbar}. As the
system is within the validity range of the thin film model of
Sec. \ref{sec:Model}, $d \ll \ell_{ex}$, the dimensionless results
thus obtained depend only weakly on the specific choices of the
dimensional parameters.

For a given pair $(\bar \kappa , \bar \delta)$ we thus choose
\begin{align}
  \label{eq:DK}
  K_u = K_d \left( 1 + {d^2 \over \ell_{ex}^2 \bar \delta^2 }
  \right), \qquad  D = {d K_d \bar \kappa \over \bar \delta}. 
\end{align}
The computational domain is discretized, using a
$2048 \times 2048 \times 2$ grid with the discretization steps
$(\Delta x, \Delta y, \Delta z)$ in the $(X,Y,Z)$ directions set to
$\Delta x = \Delta y = 0.3$ nm and $\Delta z = 1$ nm. Periodic
boundary conditions in the plane are implemented by specifying the
number of repeats in $(X, Y, Z)$ to be $(5, 5, 0)$. The exchange is
active only in the $(X,Y)$ directions.

Using the parametrization in \eqref{eq:DK}, we carried out a parameter
sweep in the $(\bar \kappa, \bar \delta)$ plane with a grid spanning
$0.05 \leq \bar \delta \leq 0.75$ with increments of
$\Delta \bar \delta = 0.025$, and $0.01 \leq \bar \kappa \leq 1.26$
with increments of $\Delta \bar \kappa = 0.025$. The maximum value of
$\bar \kappa$ is slightly below $\bar \kappa_c$, while the maximum
value of $\bar \delta$ corresponds approximately to the expected value
of skyrmion bursting threshold given by \eqref{eq:kappab2} at
$\bar \kappa = 0$. To carry out this sweep efficiently, we first
initialize the system with an antisymmetric N\'eel skyrmion
configuration for $\bar \delta = 0.05$ and $\bar \kappa = 1.1337$ on
the $\bar \kappa = \bar \kappa_2^b(\bar \delta)$ curve obtained by
running the {\tt minimize} routine with an identical Bloch
skyrmion-like initial condition of radius 5 nm in both layers to
convergence. We then march this configuration along the
$\bar \kappa = \bar \kappa_2^\mathrm{opt}(\bar \delta)$ curve by
increasing $\bar \delta$ stepwise along the grid and then running the
{\tt minimize} routine to obtain the skyrmion configurations for
$\bar \kappa = \bar \kappa_2^\mathrm{opt}(\bar \delta)$ to
convergence. We note that this curve corresponds to the optimal
stability of the skyrmion solution, and we observe, as expected, that
the numerical convergence to the skyrmion profile in the vicinity of
this optimal line is very fast. We finally use the obtained profiles
at a given $\bar \delta$ and
$\bar \kappa = \bar \kappa_2^\mathrm{opt}(\bar \delta)$ as initial
seeds to sweep down and up the values of $\bar \kappa$ along the grid
and repeat the procedure for each grid point in $\bar \delta$.
  
Next, to calculate the skyrmion collapse energy barrier defined in
\eqref{eq:dE2c} we repeated the procedure to compute the single layer
skyrmion solutions $(\m_1^\mathrm{sky1}, \m_2^\mathrm{sky1})$ with an
initial seed in the form of a skyrmion in only one layer, obtaining a
family of N\'eel skyrmions in only one layer coupled magnetostatically
to a nearly uniformly magnetized state in the other layer. We then
extracted the energy
$\bar E_2 (\m_1^\mathrm{sky1}, \m_2^\mathrm{sky1})$ of these
configurations, setting it to the last obtained value if such a
solution numerically collapsed upon decrease of $\bar \kappa$ at fixed
$\bar \delta$ in the course of the simulation.  Using \eqref{eq:dE2c},
we then calculated the collapse barrier $\Delta \bar E_2^c$ in the
parameter region in which the antisymmetric N\'eel skyrmion solution
$(\m_1^\mathrm{sky2}, \m_2^\mathrm{sky2})$ was computed.

\begin{figure*}[t]
  \centering \includegraphics[width=6.75in]{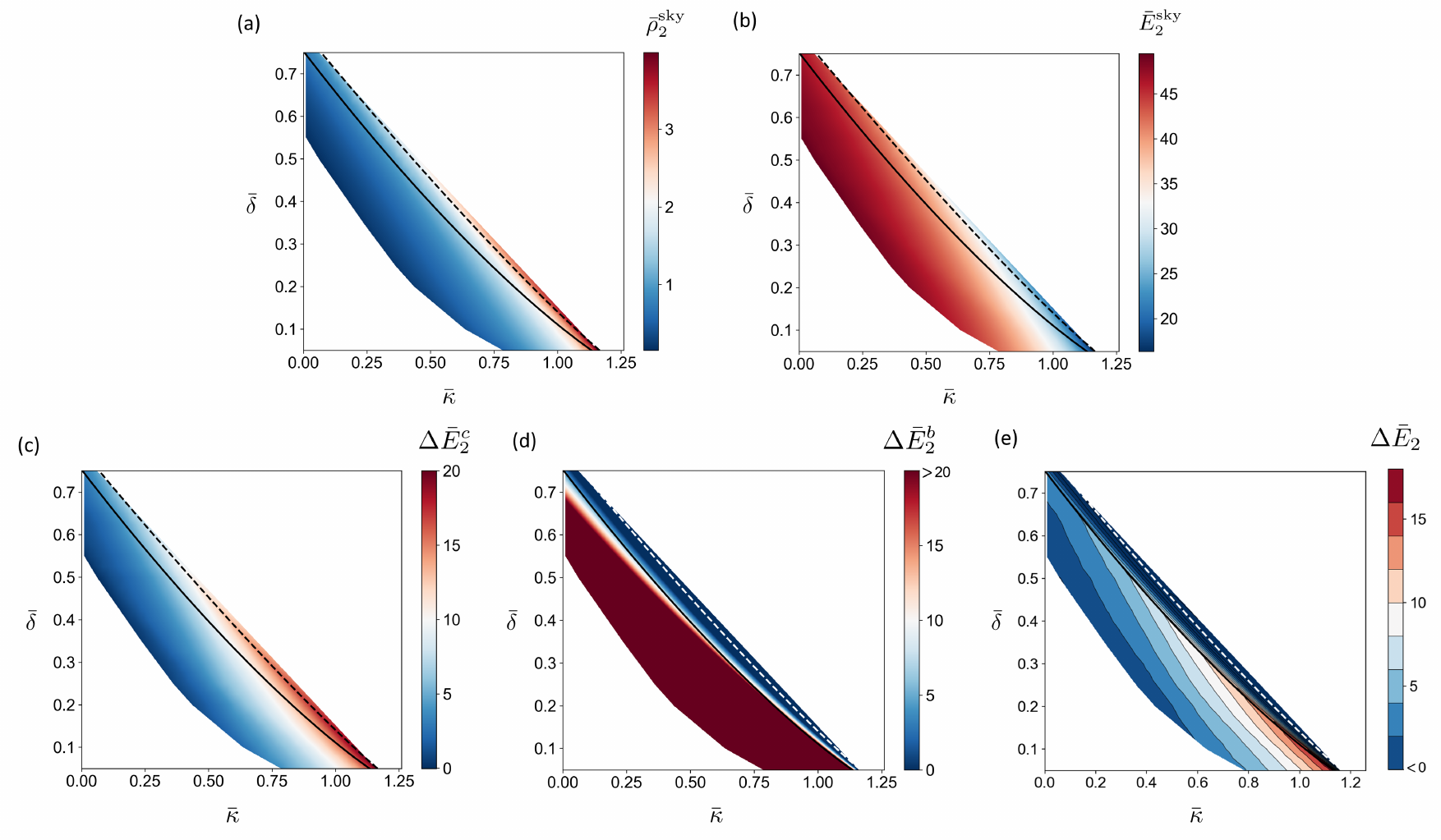}
  \caption{Summary of the numerical results obtained from {\sc Mumax3}
    simulations (see Sec.\ref{sec:numer-simul-antisymm} for details)
    of an antisymmetric ferromagnetic bilayer with $A = 20$ J/m,
    $M_s = 10^6$ A/m, $d = 1$ nm, and $K_u$ and $D$ varied in
    accordance with \eqref{eq:DK} on a $614.4 \times 614.4$ nm
    computational domain with periodic boundary conditions in the
    plane: (a) the dimensionless radius $\bar\rho_2^\mathrm{sky}$; (b)
    the dimensionless skyrmion energy $\bar E_2^\mathrm{sky}$; (c) the
    dimensionless collapse energy barrier $\Delta \bar E_2^c$; (d) the
    dimensionless bursting energy barrier $\Delta \bar E_2^b$; (e) the
    dimensionless effective energy barrier $\Delta \bar E_2$. The
    solid line shows $\bar \kappa_2^\mathrm{opt}(\bar \delta)$ and the
    dashed line shows $\bar \kappa_2^b(\bar \delta)$ governed,
    respectively, by \eqref{eq:kappa2opt} and \eqref{eq:kappab2}.}
  \label{fig:rhoE2}
\end{figure*}

The results of the above simulations in dimensionless form are
summarized in Fig. \ref{fig:rhoE2}, where the simulation data have
been linearly interpolated between the grid points.  Figure
\ref{fig:rhoE2}(a) shows the skyrmion radius
$\bar \rho_2^\mathrm{sky}$ in the units of
$L_B = \sqrt{A / (K_u - K_d)}$ (which varies from point to point)
defined as the radius of the zero level set of $m_{1,2}^\|$ from the
converged solution $(\m_1^\mathrm{sky2}, \m_2^\mathrm{sky2})$. Figure
\ref{fig:rhoE2}(b) shows the skyrmion energy $\bar E_2^\mathrm{sky}$
in the units of $Ad$ defined as the difference between the total
energy of the system and the energy of the uniformly magnetized
state. In all the simulations the resulting skyrmion profiles were
found to obey \eqref{eq:m12ansatz} and be radially symmetric. The
skyrmion solutions in Figs.  \ref{fig:rhoE2}(a) and \ref{fig:rhoE2}(b)
exhibit radii that increase with increasing $\bar \kappa$ or
$\bar\delta$, and energies that decrease with increasing $\bar \kappa$
or $\bar\delta$. In addition, the lines of constant radii and energies
extend over a wide range of $\bar \kappa$ and $\bar \delta$ values.
We also observe that for sufficiently low $\bar \kappa$ and/or
$\bar\delta$ there is no numerical solution due to numerical collapse
at fixed finite discretization in the film plane. This is due to the
numerics being unable to resolve the vanishing skyrmion radius as
$\bar\kappa, \bar\delta \to 0$, see \eqref{eq:rhob2}. Note, however,
that mathematically the skyrmion solutions continue to exist for all
sufficiently small values of $\bar \kappa$ and $\bar \delta$, as is
guaranteed by Theorem \ref{t:1}. These solutions can be obtained
numerically by reducing the discretization step $\Delta x$.  In
contrast, at high $\bar \kappa$ or $\bar\delta$ the numerical solution
bursts into a bubble occupying half of the computational domain, with
bursting occurring as predicted in
Sec. \ref{sec:skyrmion-bursting}. We exclude these solutions from the
dataset. We note a very good quantitative agreement between this
numerical bursting and the predicted busting line defined in
Eq.\eqref{eq:kappab2} and represented by a dashed line in
Figs. \ref{fig:rhoE2}(a)-(e).
  
Figure \ref{fig:rhoE2}(c) shows the plot of the bilayer collapse
barrier $\Delta \bar E_2^c$. It increases from zero at low values of
$\bar \kappa$ and $\bar \delta$ and reaches its maximum at
$\bar \kappa = \bar \kappa_2^b(\bar \delta)$ for each
$\bar\delta$. The maximum value of $\Delta \bar E_2^c$ at
$\bar \kappa = \kappa_2^b(\bar \delta)$ increases with decreasing
$\bar\delta$ and reaches the asymptotic value of
$\Delta \bar E_2^c = 8 \pi$ as $\bar \delta \to 0$ and
$\bar \kappa \to \bar \kappa_c$, corresponding to two decoupled
ferromagnetic layers.  We note that since
$\bar E_2(\m_1^\mathrm{sky1}, \m_2^\mathrm{sky1}) < 8 \pi$, the energy
of a ``zero-radius'' skyrmion in a single layer, the collapse barrier
$\Delta \bar E_2^c$ defined in \eqref{eq:dE2c} is smaller than
$\Delta \bar E_2^{c,0}$ from \eqref{eq:dE2pmc}. Therefore, by the
definition of $\bar \kappa_2^\mathrm{opt}$ [see \eqref{eq:dEequal}],
the value of the bursting energy barrier $\Delta \bar E_2^{b,0}$
predicted in \eqref{eq:dEpms} is greater than $\Delta \bar E_2^c$ for
all $\bar \kappa \leq \bar \kappa_2^\mathrm{opt}(\bar \delta)$ for a
given $\bar \delta$. Thus, one expects bursting to be active only in
the narrow range
$\bar \kappa_2^\mathrm{opt}(\bar \delta) \leq \bar \kappa \leq \bar
\kappa_2^b(\bar \delta)$ for each $\bar \delta$. This is indeed what
is observed in Fig. \ref{fig:rhoE2}(d), which shows the bursting
barrier from \eqref{eq:dEb2} as a function of $\bar \kappa$ and
$\bar \delta$, where
$\bar E_2(\m_1^\mathrm{sad}, \m_2^\mathrm{sad}) \simeq 2 \bar
E^\pm(\m_{\bar \rho_2^\mathrm{sad}})$ is the approximate saddle point
energy from \eqref{eq:E2sad} and
$\bar E_2(\m_1^\mathrm{sky2}, \m_2^\mathrm{sky2})$ is the skyrmion
energy obtained numerically, all in the units of $Ad$ [we set
$\Delta \bar E_2^b = 0$ when the above definition yields a negative
value, indicating the breakdown of validity of \eqref{eq:E2sad}]. As
expected, the bursting barrier exceeds the value of
$\Delta \bar E_2^c$ to the left of the
$\bar \kappa = \bar \kappa_2^\mathrm{opt}(\bar \delta)$ curve, but
very quickly drops off to zero as $\bar \kappa$ approaches
$\bar \kappa_2^b(\bar \delta)$. Thus, the optimal thermal stability of
a skyrmion indeed occurs at
$\bar \kappa \simeq \bar \kappa_2^\mathrm{opt}(\bar \delta)$ for every
fixed $\bar \delta$. This is illustrated in Fig. \ref{fig:rhoE2}(e),
which shows the effective energy barrier
$\Delta \bar E_2 = \min( \Delta \bar E_2^c, \Delta \bar E_2^b)$. In
Fig. \ref{fig:rhoE2}(e), we see that below the curve
$\bar \kappa = \bar \kappa_2^\mathrm{opt}(\bar\delta)$ the lines of
equal effective energy barrier $\Delta \bar E_2$ again cover a wide
range of $\bar \kappa$ and $\bar \delta$ values. This highlights the
existence of two degrees of freedom, $\bar \kappa$ and $\bar \delta$,
for the optimization of skyrmion stability in the antisymmetric
bilayer system. It is a clear confirmation that in this system the
stray field plays an important stabilizing role via the surface-volume
charge interactions, cooperating with the DMI, as could be inferred by
looking at the skyrmion energy in \eqref{eq:E2b0}.

  \section{Comparison with synthetic antiferromagnets and
    ferromagnetic monolayers}
\label{sec:comp}

In order to emphasize the potential advantages of the antisymmetric
bilayer over other systems, we compare it with the typical layer
stacks for skyrmion applications, represented in
Figs.~\ref{fig:anti}(b) and \ref{fig:anti}(c), namely, SAF and
ferromagnetic monolayers. We start by presenting the models and
comparatively discussing the numerical skyrmion solutions for the case
of SAF in Sec.  \ref{sec:synth-antiferro-bilay} and ferromagnetic
monolayers in Sec.  \ref{sec:monolayers}, respectively. Finally, in
Section \ref{sec:comparison} we apply our predictions to the family of
prototypical spintronic materials, the Pt/Co/AlO$_x$ stacks
\cite{dahmane08,belmeguenai15,krishnia25}. The radii and effective
energy barriers for skyrmions in an antisymmetric bilayer, SAF and a
monolayer constructed with this material as a building block are given
as a function of dimensional thickness and interfacial DMI, and then
compared to one another.

\subsection{Synthetic antiferromagnet}
\label{sec:synth-antiferro-bilay}

As was already mentioned, SAF multilayers are among the most promising
systems to achieve room temperature stable magnetic skyrmions. They
consist of a basic unit in which two identical ferromagnetic
monolayers with the DMI of the same sign in each layer are strongly
antiferromagnetically coupled, leading effectively to a cancellation
of the non-local magnetostatic interactions. Such an additional
interlayer coupling may be easily incorporated into our modeling
framework, amounting to an additional term in the micromagnetic energy
in \eqref{eq:E3dM} \cite{nogues05,dubicki:phd}:
\begin{align}
  \label{eq:EJ}
  \mathcal E_J
  & (\mathbf M) = \mathcal E(\mathbf M) 
    \notag \\
  & - \frac{J}{M_s^2} \sum_{n=1}^{N-1} 
    \int_{\R^2}\Big( \mathbf M_n^+ \cdot \mathbf M_{n+1}^- 
    - \mathbf M_n^\infty
    \cdot \mathbf M_{n+1}^\infty \Big) d^2 r,
\end{align}
where $J$ is the interlayer exchange coupling strength (in J/m$^2$),
$\mathbf M_n^\pm$ are the magnetization vectors evaluated on
$\partial \Omega_n^\pm$, and
$\mathbf M_n^\infty = \mathbf M_n(\infty)$, with the last term
introduced to avoid divergent integrals. A non-dimensionalization as
in \eqref{eq:E3dm}, followed by the asymptotic expansion in
$\delta \ll 1$ then results in
$\mathcal E_J(\mathbf M)/(Ad) \simeq E_N^\lambda(\m_1, \ldots, \m_N)$,
where
\begin{align}
  \label{eq:Elam}
  & E_N^\lambda
    (\m_1, \ldots, \m_N) 
    = E_N (\m_1, \ldots, \m_N) \notag \\
  & \ + \frac{\lambda}{2}
    \sum_{n=1}^{N-1} 
    \int_{\R^2}  \Big( |\m_n - \m_{n+1} |^2
    - |\m_n^\infty
    - \m_{n+1}^\infty|^2 \Big) d^2 r,
\end{align}
with $\lambda = J / (d K_d)$ the dimensionless interlayer coupling
strength and $\m_n^\infty = \m_n(\infty)$. It is easy to see that if
the interlayer exchange coupling strength is similar to the intralayer
exchange, we have $\lambda \sim 1/\delta^2 \gg 1$.

\begin{figure}[t]
  \centering
  \includegraphics[width=3.15in]{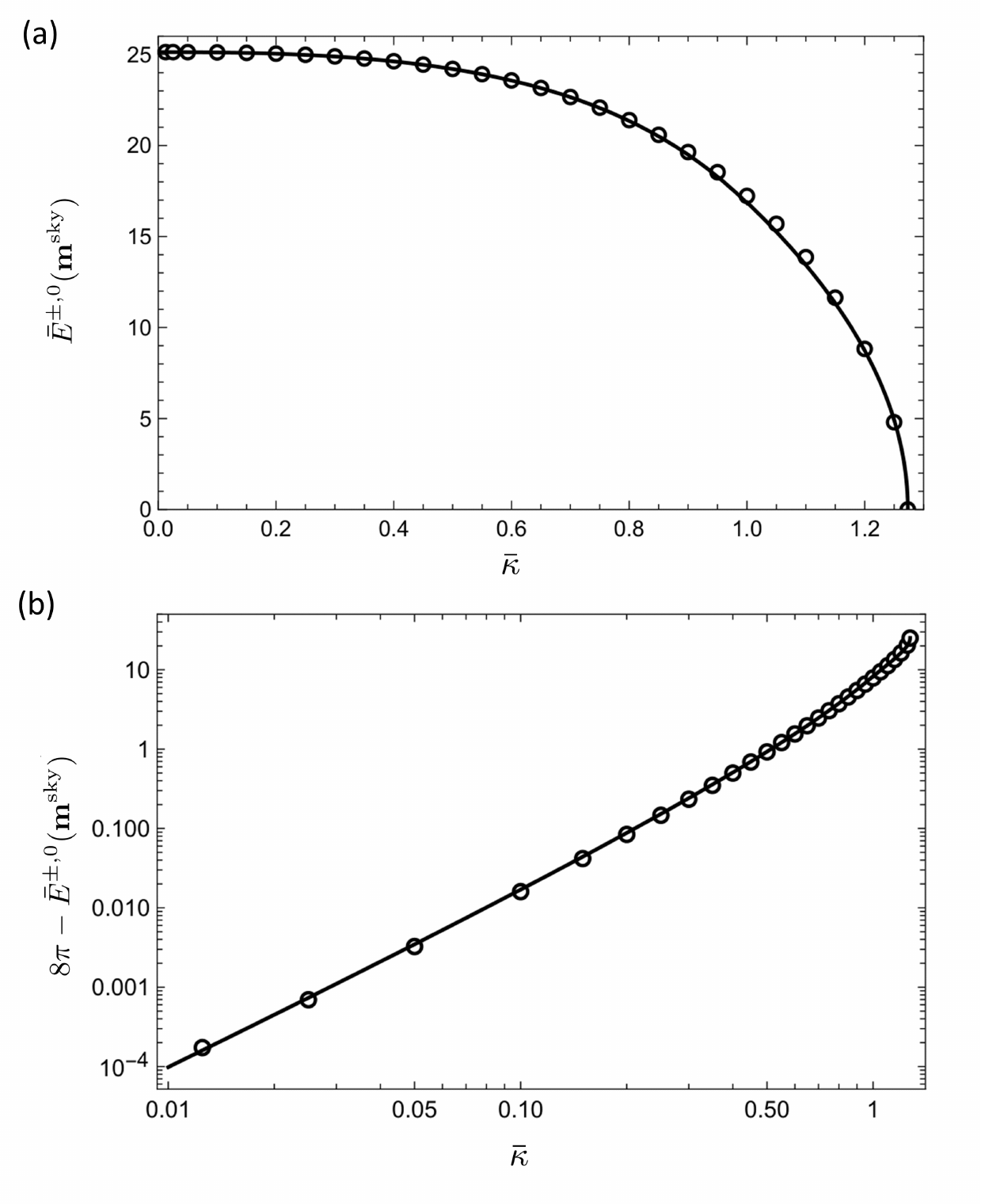}   
  \caption{Comparison of the skyrmion energy obtained numerically
    (open circles) with the prediction of \eqref{eq:dEbc0} (solid
    lines): (a) the graph of $\bar E^{\pm,0}(\m^\mathrm{sky})$
    vs. $\bar\kappa$; (b) the graph of
    $8 \pi - \bar E^{\pm,0}(\m^\mathrm{sky})$ vs. $\bar\kappa$. }
  \label{fig:Efit}
\end{figure}

For SAF in the form of a bilayer, we have $N = 2$, $\lambda < 0$ and
$|\lambda| \gg 1$, which forces antiferromagnetic order:
$\m_1 \simeq \m$ and $\m_2 \simeq -\m$ for some
$\m : \R^2 \to \mathbb S^2$. As a result, to the leading order in
$|\lambda| \gg 1$ we have, after a further rescaling of space,
$\bar E_2^\lambda \simeq 2 \bar E^{\pm,0}$, where $\bar E^{\pm,0}$ is
given by \eqref{eq:E2b0} with $\bar \delta = 0$. It corresponds to the
classical two-dimensional micromagnetic energy without non-local
effects that has been extensively studied
\cite{bogdanov89,bogdanov89a,rohart13,bms:prb20,komineas20,
  komineas21,gustafson21,bms:pnas22,juge22, komineas23}. The
asymptotic behavior of the skyrmion energy as $\bar \kappa \to 0$ was
established in \cite{bms:prb20} (see also \cite{komineas20,bms:arma21,
  gustafson21}), while the asymptotic behavior for
$\bar\kappa \to \bar\kappa_c = 4/\pi$ was established in
\cite{komineas21} (see also \cite{rohart13}). Various ad-hoc formulas
for the skyrmion energy have been proposed
\cite{bms:pnas22,juge22,komineas23}. These are able to produce
reasonable approximations for the skyrmion energy for certain ranges
of $\bar \kappa$, but fail to provide a uniformly accurate
approximation for the entire expected range
$0 < \bar \kappa < \bar \kappa_c$ of skyrmion existence. Here we
rectify this issue by providing a formula for the skyrmion collapse
energy barrier
$\Delta \bar E_2^{c,0} = 16 \pi - 2 \bar E^{\pm,0}(\m^\mathrm{sky})$
that agrees within a relative error of 6\% for the entire range of
$\bar\kappa$ with the one calculated from the ``exact'' skyrmion
energy obtained from the numerical solution of the Euler-Lagrange
equation associated with $\bar E^{\pm,0}(\m^\mathrm{sky})$ by the
shooting method. The formula reads
\begin{align}
  \label{eq:dEbc0}
  \bar E^{\pm,0}(\m^\mathrm{sky}) \simeq 
  \begin{cases}
    8 \pi - {2 \pi \bar \kappa^2 \over \ln \left[ {\alpha \over
          \bar\kappa} \ln \left( {16 \alpha \over \bar\kappa} \right)
      \right]} & \ 0 <
    \bar \kappa \leq 1, \vspace{1mm} \\
    8 \pi \beta \sqrt{2(\bar \kappa_c - \bar \kappa) / \bar \kappa_c}
    & \ 1 < \bar \kappa < \bar \kappa_c,
  \end{cases}
\end{align}
where $\alpha = 4 e^{-1-\gamma} \approx 0.8262$ and $\beta =
1.025$. The small $\bar\kappa$ formula is obtained from the result in
\cite{bms:prb20} by keeping the
$O(\ln |\ln \bar\kappa| / |\ln \bar\kappa|)$ term in the expansion and
fitting the constant in the inner logarithm to the numerical data,
while the large $\bar\kappa$ formula is obtained from the expansions
in \cite{rohart13,komineas21}, with the coefficient of the square root
slightly adjusted to fit the numerics. A comparison of the skyrmion
energy predicted by \eqref{eq:dEbc0} and the one obtained numerically
is presented in Fig. \ref{fig:Efit}. The formula also gives the
skyrmion energy $\bar E^{\pm,0}(\m^\mathrm{sky})$ with a uniform
relative error of at most $3$\% for all values of $\bar\kappa$.

We also mention an empirical formula obtained for the skyrmion radius
$\bar \rho_{2,0}^\mathrm{sky}$ in the units of $L_B$ for synthetic
antiferromagnetic bilayers obtained in \cite{pham24}:
\begin{align}
  \label{eq:rhoSAF}
  \bar \rho_{2,0}^\mathrm{sky} = {a \bar \kappa^2 \over \bar \kappa_c
  \sqrt{\bar \kappa_c^2 - \bar \kappa^2}},
\end{align}
where $a = 1.35$, which gives an approximation for
$\bar \rho_{2,0}^\mathrm{sky}$ to within a 5\% relative error in the
range $0.35 \leq \bar \kappa \leq 1.15$, and within a 25\% relative
error in the range $0.2 \leq \bar \kappa \leq 1.2$. Notice that this
formula fails to reproduce the asymptotic behavior of
$\bar \rho_{2,0}^\mathrm{sky}$ as a function of $\bar \kappa$ for
$\bar \kappa \to 0$ or $\bar \kappa \to \bar \kappa_c$
\cite{bms:prb20,komineas20,komineas21}.

To summarize, in a SAF, the skyrmion solutions have their energies and
radii independent of $\bar \delta$ due to the cancellation of the
non-local magnetostatic interactions. Another consequence of the
vanishing stray field is the absence of bursting and the persistence
of the skyrmion solution up to $\bar \kappa_c$. However, as can be
seen from Fig. \ref{fig:Efit}(b), the skyrmion collapse energy is
dropping quickly away from $\bar \kappa_c$ .

\begin{figure*}[t]
  \centering
  \includegraphics[width=6.75in]{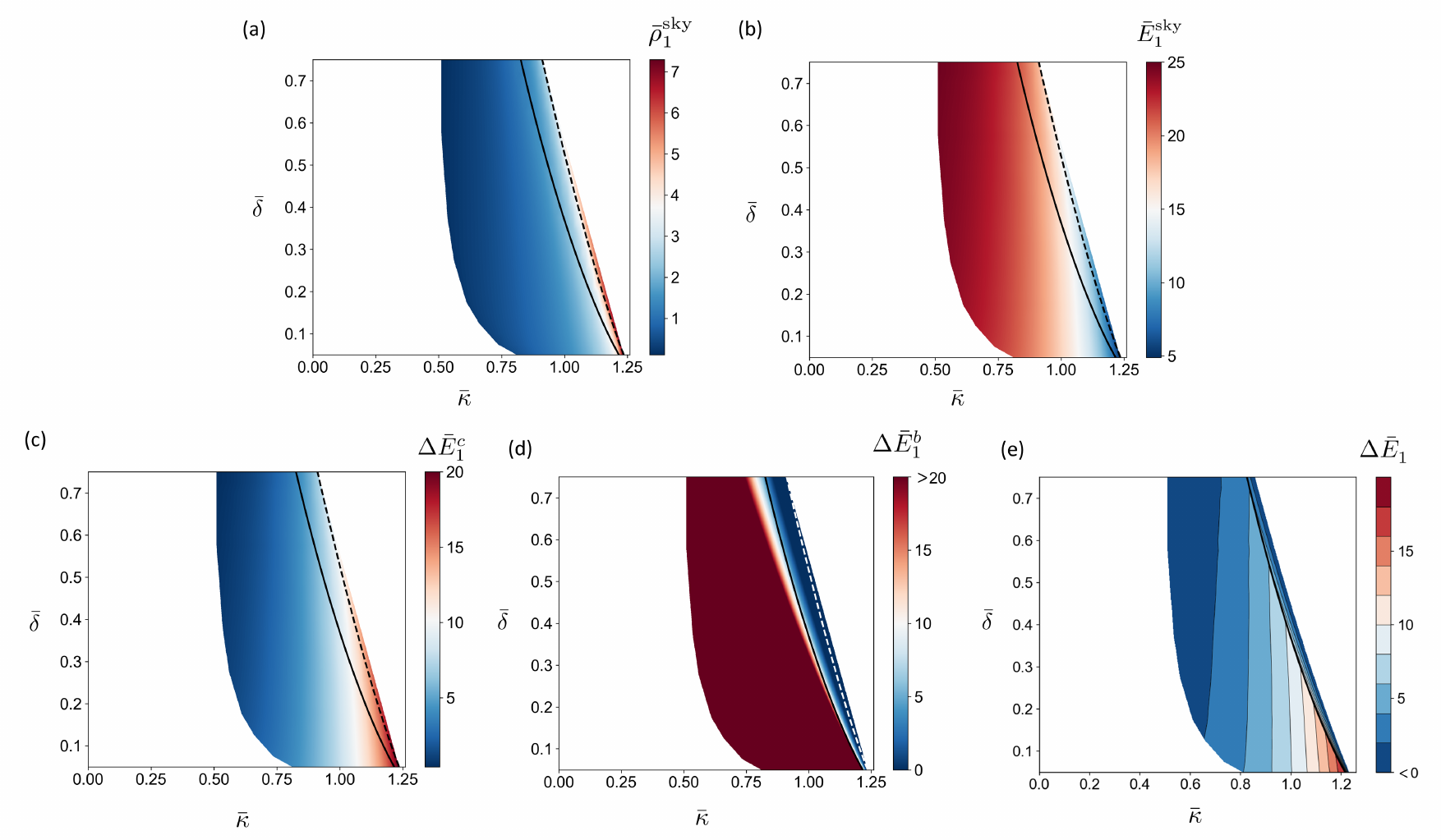}
  \caption{Summary of the numerical results obtained from {\sc Mumax3}
    simulations (see Sec. \ref{sec:monolayers} for details) of a
    ferromagnetic monolayer with the same parameters as in
    Fig. \ref{fig:rhoE2}: (a) the dimensionless radius
    $\bar\rho_1^\mathrm{sky}$; (b) the dimensionless skyrmion energy
    $\bar E_1^\mathrm{sky}$; (c) the dimensionless collapse energy
    barrier $\Delta \bar E_1^c$; (d) the dimensionless bursting energy
    barrier $\Delta \bar E_1^b$; (e) the dimensionless effective
    energy barrier $\Delta \bar E_1$. The solid line shows
    $\bar \kappa_1^\mathrm{opt}(\bar \delta)$, while the dashed line
    shows $\bar \kappa_1^b(\bar \delta)$, governed, respectively, by
    \eqref{eq:kappa1opt} and \eqref{eq:kb1}.}
  \label{fig:rhoE1}
\end{figure*}

\subsection{Ferromagnetic monolayers}
\label{sec:monolayers}

We finally discuss the case of ferromagnetic monolayers, which also
applies to symmetric ferromagnetic multilayers with strong
ferromagnetic exchange coupling, $\lambda \gg 1$ in \eqref{eq:Elam},
upon taking $d$ to be the sum of all the ferromagnetic layer
thicknesses. We start by noting that for ultrathin monolayers, the
skyrmion solutions in the low $\bar \kappa $ and $\bar \delta$ regime
were characterized in our previous work by rigorous asymptotic
analysis \cite{bms:prb20,bms:arma21,bfbsm:prb23}. Skyrmions exhibit
compact profiles close to the Belavin-Polyakov profiles and a
reorientation from the Néel rotation at $\bar \delta =0$ to the Bloch
rotation at $\bar \kappa =0$. For larger $\bar \kappa $ and
$\bar \delta$, beyond the compact skyrmion limit, Büttner et
al. \cite{buttner18} calculated numerically the skyrmion
characteristics using a 360\textdegree-wall ansatz, without computing
the skyrmion bursting threshold. This threshold was discussed by
Bernand-Mantel et al. \cite{bernand-mantel18} and characterized by a
numerical study of an analytical model based on the thin wall
approximation valid for layer thicknesses much larger than the Bloch
wall width, $d \gg L_B$ \cite{kaplan93}. The model and the numerical
solutions presented here complement those previous findings in the
ultrathin film regime, $d \lesssim L_B$, and not too small DMI
strengths, which forces the Néel rotation.  As the analysis follows
closely the steps of Sec. \ref{sec:abl}, we only provide the necessary
modifications and then present the numerical results obtained for the
monolayer.

As was shown in \cite{bms:prb20}, skyrmion solutions exist for all
$0 < 2 \bar \kappa + \bar \delta \leq \sqrt{2}$, with the profiles of
N\'eel type for all $\bar \delta \leq {32 \over 3 \pi^2} \bar \kappa$
and all sufficiently small $\bar \kappa$. For larger values of
$\bar \kappa$, one can carry out the same analysis as in
Sec. \ref{sec:skyrmion-bursting} to obtain the following expression
for the energy of a bubble-like profile in \eqref{eq:bubble}:
\begin{align}
  \label{eq:E1rho}
  & \bar E_1(\m_{\bar \rho}) \simeq 2 \pi \bar \rho \left[4 - \pi \bar
    \kappa + {2 \bar \delta \ln 2 \over \pi} \right] + {4 \pi \over
    \bar \rho} \notag \\ 
  & \quad - 4 \bar \delta  {\bar\rho} \ln
    \left(\frac{8 {\bar\rho}}{\pi } e^{\gamma - 1} \right),
\end{align}
to the leading order in ${\bar\rho} \gg 1$, where we accounted for an
additional energy penalty arising from the volume-volume charge
interactions \cite{tarasenko98,thiaville12}. Here the profile is
expected to retain its N\'eel character in a much broader range
$\bar \delta \leq {\pi^2 \over 4 \ln 2} \bar \kappa$ corresponding to
the threshold for a 1D domain wall \cite{thiaville12}.

As in the case of antisymmetric bilayers, the expression in
\eqref{eq:E1rho} predicts the existence of a critical value of $\bar
\kappa = \bar \kappa_1^b$ given by
\begin{align}
  \label{eq:kb1}
  \bar \kappa_1^b \simeq {4 \over \pi} - {\bar \delta \over \pi^2}
  \left[ \ln \left( {32 \over \pi \bar \delta } \right) + 2 \gamma + 1
  \right], 
\end{align}
such that two critical points of $\bar E_1(\m_{\bar \rho})$ exist for
all $\bar \kappa < \bar \kappa_1^b(\bar \delta)$, while the energy is
monotonically decreasing for all
$\bar \kappa > \bar \kappa_1^b(\bar \delta)$, indicating skyrmion
bursting at $\bar \kappa = \bar \kappa_1^b$. The formula is expected
to be valid for $\bar \delta \lesssim 1$.

The expression in \eqref{eq:E1rho} may also be used to predict the
energy barrier against bursting. When $\bar \kappa$ is not too close
to $\bar \kappa_c$, the saddle point radius $\bar \rho_1^\mathrm{sad}$
may be obtained by dropping the $4 \pi / \bar \rho$ term from
\eqref{eq:E1rho} and finding the critical point of the energy
\begin{align}
  \label{eq:rho1sad}
  \bar \rho_1^\mathrm{sad} \simeq {\pi \over 4} \exp \left( {2 \pi
  \over \bar \delta} - {\pi^2 \bar \kappa \over 2 \bar \delta} - \gamma
  \right). 
\end{align}
The associated bursting energy barrier is
$\Delta \bar E_1^b = \bar E_1(\m_{\bar\rho_1^\mathrm{sad}}) - \bar
E_1(\m^\mathrm{sky})$, where
\begin{align}
  \label{eq:E1sad}
  \bar E_1(\m_{\bar\rho_1^\mathrm{sad}}) \simeq \pi \bar \delta \exp \left( {2
  \pi \over \bar \delta} - {\pi^2 \bar \kappa \over 2 \bar \delta} - \gamma
  \right), 
\end{align}
and $\bar E_1(\m^\mathrm{sky})$ is the energy of the skyrmion
solution. The collapse barrier, on the other hand, is
$\Delta \bar E_1^c = 8 \pi - \bar E_1(\m^\mathrm{sky})$
\cite{buttner18,bernand-mantel18,heil19,bms:prb20,bms:pnas22}, so
equating the two we obtain the predicted optimal value
$\bar \kappa_1^\mathrm{opt}$ at which the skyrmion lifetime is
maximized at fixed $\bar \delta \lesssim 1$:
\begin{align}
  \label{eq:kappa1opt}
  \bar \kappa_1^\mathrm{opt} \simeq {4 \over \pi} - {2 \bar \delta
  \over \pi^2} \ln \left( { 8 e^\gamma \over \bar \delta} \right). 
\end{align}
We finally note that the relation in \eqref{eq:rsLopt} also holds in
the case of monolayers for $\bar\delta \ll 1$ and $\bar\kappa$ not too
close to $\bar \kappa_c$ (cf. \cite{bsm:prb25}).

We conclude this section by carrying out the simulations for
monolayers that are analogous to those of
Sec. \ref{sec:numer-simul-antisymm}. All the simulations use the same
parameters as in Sec. \ref{sec:numer-simul-antisymm}, except the
computational grid used is $2048 \times 2048 \times 1$. The computed
dimensionless skyrmion radius $\bar \rho_1^\mathrm{sky}$ and energy
$\bar E_1(\m^\mathrm{sky})$ as functions of $\bar \kappa$ and
$\bar \delta$ are presented in Figs. \ref{fig:rhoE1}(a) and
\ref{fig:rhoE1}(b). 
In turn, the dimensionless collapse energy barrier
$\Delta \bar E_1^c = 8 \pi - \bar E_1(\m^\mathrm{sky})$ is plotted in
Fig. \ref{fig:rhoE1}(c), while the bursting energy barrier
$\Delta \bar E_1^b = \bar E_1(\m^\mathrm{sad}) - \bar
E_1(\m^\mathrm{sky})$, where
$\bar E_1(\m^\mathrm{sad}) \simeq \bar E_1(\m_{\bar
  \rho_1^\mathrm{sad}})$ is given by \eqref{eq:E1sad} is plotted in
Fig. \ref{fig:rhoE1}(d). The effective energy barrier
$\Delta \bar E_1 = \min(\Delta \bar E_1^c , \Delta \bar E_1^b)$ is
shown in Fig. \ref{fig:rhoE1}(e).

Contrary to the case of antisymmetric bilayers, the skyrmion solution
radii and collapse barriers in Figs. \ref{fig:rhoE1}(a) and
\ref{fig:rhoE1}(c) show mostly an increase with increasing
$\bar \kappa$, while their variation with $\bar \delta$ is weak.
Again, the absence of the numerical skyrmion solution for low
$\bar \kappa $ and $\bar \delta$ is due to the fixed finite
discretization in the film plane that is unable to resolve the
skyrmion of vanishingly low radius and leads to a numerical collapse
(see also the discussion in Section
\ref{sec:numer-simul-antisymm}). As in the case of the antisymmetric
bilayer in Sec. \ref{sec:numer-simul-antisymm}, for large values of
$\bar \kappa $ and $\bar \delta $ we observe bursting of the skyrmion,
which shows a very good agreement with the predicted bursting
threshold $\bar \kappa_1^b(\bar \delta)$ from \eqref{eq:kb1},
represented by a dashed line in Fig. \ref{fig:rhoE1}. Below the
optimal stability threshold $\bar \kappa_1^\mathrm{opt}(\bar \delta)$
(solid line), the effective energy barrier shown in
Fig. \ref{fig:rhoE1}(e), shows a weak dependence on $\bar \delta$. In
fact, we observe that, contrary to the antisymmetric bilayer case, the
level lines of sufficiently high effective energy barrier are covering
very narrow ranges of $\bar \kappa$ values.

\subsection{Comparison of skyrmion solutions for a typical
  ferromagnetic material}
\label{sec:comparison}

We now demonstrate how the obtained results can be used to predict
skyrmion stability against thermal noise, using a typical family of
ferromagnetic materials, the Pt/Co/AlO$_x$ system, as a building block
to construct the three types of stacking in Fig. \ref{fig:anti}.

We begin by pointing out the fundamental energy scale associated with
the thermally activated processes in ultra-thin film micromagnetics,
namely, the energy of a ``zero-radius skyrmion'' in a monolayer,
$\Delta \mathcal E_0 = 8 \pi A d$, corresponding to the energy of the
minimizers of the exchange energy alone in the form of
Belavin-Polyakov profiles \cite{belavin75}. With the parameters that
we chose to represent a characteristic ferromagnetic Pt/Co/AlO$_x$
layer in Sec. \ref{sec:numer-simul-antisymm}, $A=20$ pJ/m and $d$=1
nm, we have $\Delta \mathcal E_0 \approx 120 k_BT_{RT}$, where
$T_{RT}=300K$ is the room temperature. It sets an upper bound for the
collapse energy barrier.

As this fundamental energy scale $\Delta \mathcal E_0$ increases
with the film thickness, it is natural to consider stacking multiple
ferromagnetic layers to enhance the effective thickness without
decreasing the DMI \cite{buttner18}.  However, as we show in the
present work, when $d$ is increased, the stray field effects get
stronger and the skyrmion solution disappears above a critical value
of the thickness corresponding to the bursting threshold determined by
\eqref{eq:kappab2} or \eqref{eq:kb1} in the cases of antisymmetric
bilayers or monolayers, respectively.  An additional difficulty is the
dependence of the micromagnetic parameters on the film thickness
\cite{belmeguenai15,krishnia25}, so predictions about skyrmion
existence and stability require the knowledge of all these specific
dependences.

\begin{figure*}[t]
  \centering
  \includegraphics[width=6.75in]{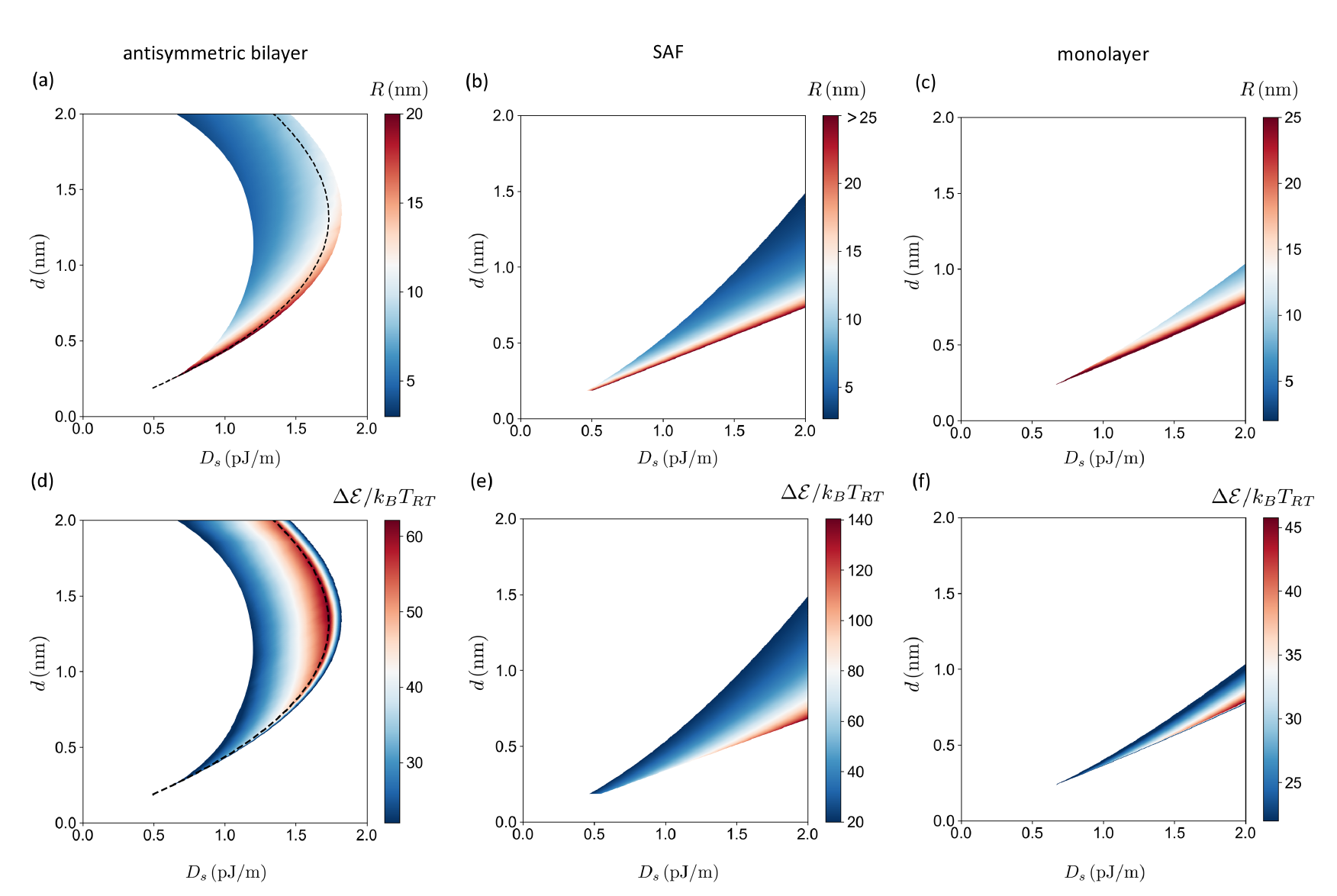}
  \caption{Skyrmion radius $R$ (a)-(c) and effective energy barrier
    $\Delta \mathcal E$ in the units of $k_BT_{RT}$ (d)-(f), where we
    excluded skyrmion solutions with
    $\Delta \mathcal E < 20 k_BT_{RT}$, as functions of the
    interfacial DMI strength $D_s$ and the single ferromagnetic layer
    thickness $d$ for: (a,d) antisymmetric ferromagnetic bilayer;
    (b,e) SAF ; (c,f) ferromagnetic monolayer. In (a) and (d):
    dimensional versions of the skyrmion radius
    $R=\bar \rho^\mathrm{sky}_2 L_B$ and the effective energy barrier
    $\Delta \mathcal E = A d \Delta \bar E_2$ are dimensional replots
    of the data from the skyrmion solutions presented in
    Figs. \ref{fig:rhoE2}(a) and \ref{fig:rhoE2}(e) from
    Sec. \ref{sec:numer-simul-antisymm}. In (b) and (e): dimensional
    versions of the skyrmion radius
    $R= \bar \rho_{2,0}^\mathrm{sky} L_B$, where
    $ \bar \rho_{2,0}^\mathrm{sky}$ is given by \eqref{eq:rhoSAF}, and
    effective energy barrier
    $\Delta \mathcal E = A d \Delta \bar E_2^{c,0}$ where
    $\Delta \bar E_2^{c,0}$ is given by \eqref{eq:dEbc0}. In (c) and
    (f): the dimensional version of the skyrmion radius
    $R=\bar \rho^\mathrm{sky}_1 L_B$ and the barrier
    $\Delta \mathcal E = A d \Delta \bar E_1$ are dimensional replots
    of the data from the skyrmion solutions presented in
    Figs. \ref{fig:rhoE1}(a) and \ref{fig:rhoE1}(e) from
    Sec. \ref{sec:monolayers}.  Other parameters in this figure are:
    $A = 20$ pJ/m, $M_s = 1$ MA/m, $K_u = 0.9$ MJ/m$^3$ and $K_s = 0$
    J/m$^2$. The dashed line in (a) and (d) shows
    $D_{s,\mathrm{opt}}^\mathrm{bi}$ vs. $d$ from \eqref{eq:Dopt2}.}
  \label{fig:DE}
\end{figure*}

To circumvent this difficulty, in the following we make a simplifying
assumption that the ferromagnet exhibits a bulk perpendicular
magnetocrystalline anisotropy $K_u > K_d$ independent of $d$ and a
negligible interfacial anisotropy $K_s = 0$ (in J/m$^2$). While in
principle the magnetocrystalline anisotropy is of interfacial origin
in the Pt/Co/AlO$_x$ system, this choice enables us to highlight the
effect of the modulation of the stray field via the dimensional
thickness change independently of the anisotropy change. This is also
experimentally justified, considering the recent finding in
Pt/Co/AlO$_x$ of a strong anisotropy tuning, at fixed Co thickness,
via a thickness increase of the capping Al layer, which enables an
independent tuning of Co thickness and anisotropy in this system
\cite{krishnia25}.  We also assume that the DMI is of interfacial
origin, which corresponds to a DMI constant per unit volume
$D = D_s / d$ (in the units of J/m$^2$) with the interfacial DMI
strength $D_s$ (in the units of J/m) independent of $d$.

With the above assumptions, we study the resulting dependence of the
energy barriers governing skyrmion stability on the parameters and, in
particular, on the film thickness as the most easily adjustable
material parameter. It is clear that since the barrier height vanishes
as $d \to 0$ due to the dimensional $Ad$ factor, and since the
solution disappears for $d$ above a critical value depending on the
rest of the parameters, there must exist a set of parameters for which
the energy barrier attains its maximum value.

The predicted dependences of the skyrmion radius $R$ and the effective
energy barrier $\Delta \mathcal E$ on $D_s$ and $d$ for an
antisymmetric bilayer, a SAF and a ferromagnetic monolayer with
$A = 20$ pJ/m, $M_s = 1$ MA/m and $K_u = 0.9$ MJ/m$^3$, in the
physically realizable range $0 < D_s < D_s^\mathrm{max}$ with
$D_s^\mathrm{max} \simeq 2$ pJ/m \cite{belmeguenai15,krishnia25}, are
presented in Fig. \ref{fig:DE}. These graphs are obtained by
replotting the dimensional skyrmion radii $R$ and the effective energy
barriers $\Delta \mathcal E$ calculated from the dimensionless radii
$\bar \rho$ and effective energy barriers $\Delta \bar E$ obtained in
the previous sections as functions of $\bar \kappa$ and $\bar \delta$
expressed in terms of the interfacial DMI strength $D_s$ and the
single ferromagnetic layer thickness $d$. As our aim is to focus on
room temperature stability, we only present the skyrmion solutions
with effective barriers $\Delta \mathcal E > 20k_BT_{RT}$ in
Fig. \ref{fig:DE}.

\begin{figure}[t]
  \centering
  \includegraphics[width=3.5in]{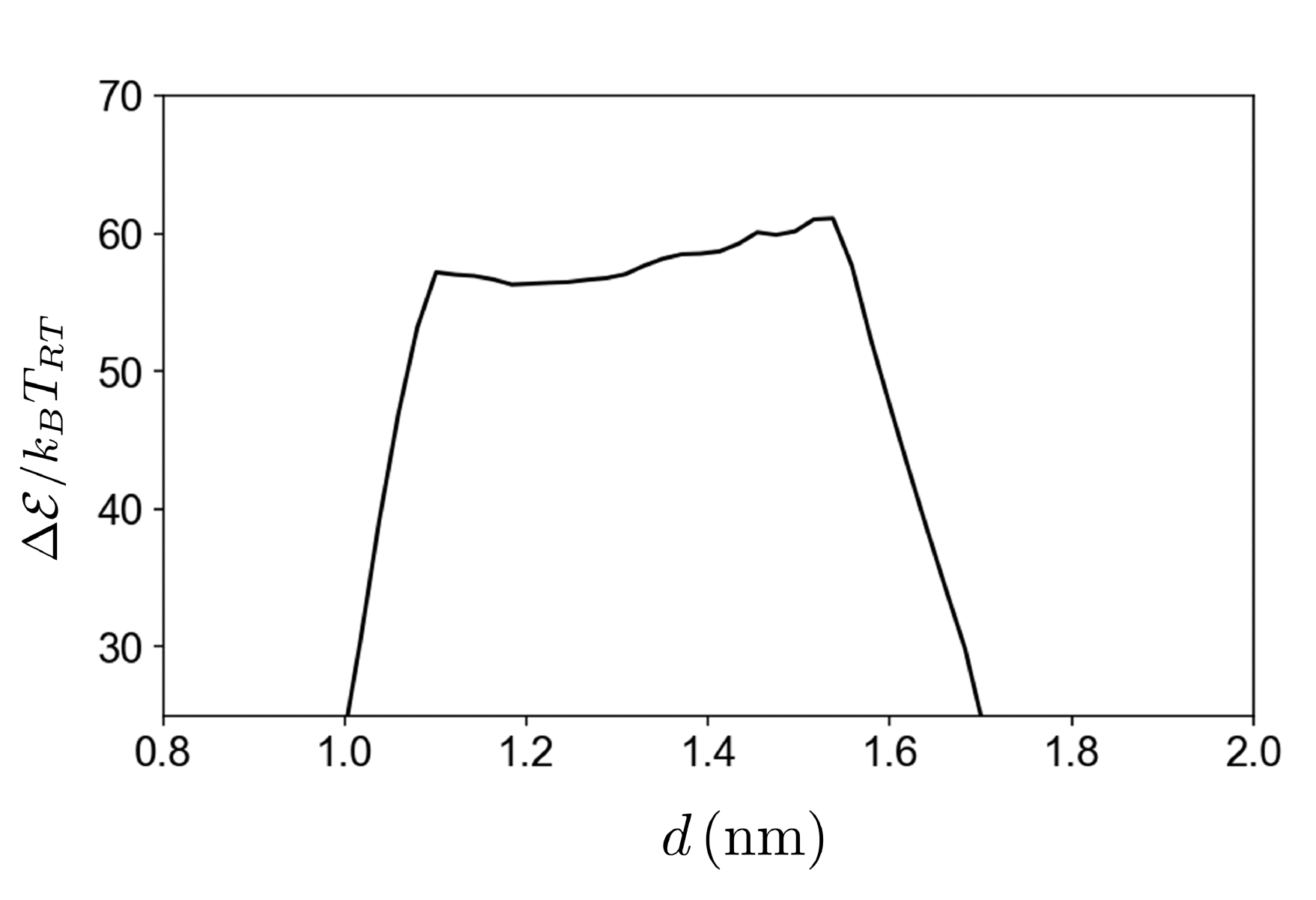}
  \caption{Effective energy barrier $\Delta \mathcal E$ in the units
    of $k_BT_{RT}$ as a function of $d$ obtained from
    Fig. \ref{fig:DE}(e) for the antisymmetric ferromagnetic bilayer
    with fixed $D_s = 1.7$ pJ/m.}
  \label{fig:DEbiDs}
\end{figure}

For antisymmetric bilayers in Figs. \ref{fig:DE}(a) and
\ref{fig:DE}(d), we present the radii $R$ and the effective energy
barriers $\Delta \mathcal E$ of skyrmions as functions of $D_s$ and
$d$ (see figure caption for details). The skyrmion solutions are
extending across a large crescent-shaped region.  The maximum
effective energy barrier in the crescent is reaching
$\simeq 62 k_BT_{RT}$ at $D_s \simeq 1.74$ pJ/m and $d \simeq 1.4$ nm,
and corresponds to a skyrmion solution of radius $R \simeq 11$ nm. We
note that the optimal skyrmion stability occurs, for every thickness
$d$, in the close vicinity of $D_s = D_{s,\mathrm{opt}}^\mathrm{bi}$,
represented by a dashed line in Figs. \ref{fig:DE}(a) and
\ref{fig:DE}(d), that is obtained from \eqref{eq:kappa2opt}:
\begin{align}
  \label{eq:Dopt2}
  & D_{s,\mathrm{opt}}^\mathrm{bi} \simeq {4 d \over \pi} \sqrt{A (K_u
    -  K_d)} \notag 
  \\  
  & \quad - {4 d^2 K_d
    \over \pi^2} \left[ \ln \left( {8 \sqrt{A (K_u -
    K_d)} \over d K_d } \right) + \gamma + {\pi^2 \over 8}
    \right].
\end{align}
The barrier height remains uniformly higher than $50 k_BT_{RT}$ in a
broad range of thicknesses and in a certain range of $D_s$, with the
radii varying from 8 nm to 15 nm. For example, the cross-section of
the plot in Fig. \ref{fig:DE}(d) at $D_s = 1.7$ pJ/m is shown in
Fig. \ref{fig:DEbiDs}, indicating this large range of thermal
stability against variatons of $d$. These results identify the
antisymmetric bilayer as a promising platform for 10 nm radius
zero-field skyrmions with lifetimes compatible with information
technology applications.

For SAF in the form of a bilayer, in Figs. \ref{fig:DE}(b) and
\ref{fig:DE}(e) we display the radii $R$ and the collapse energy
barriers $\Delta \mathcal E$ as functions of $D_s$ and $d$ (see figure
caption for details). Here the region of relative stability
($\Delta \mathcal E \geq 20 k_BT_{RT}$) has the form of a wedge, in
which the barrier height can reach its theoretical maximum of
$16 \pi A d$ as $D = D_s d$ approaches
$D_c = {4 \over \pi} \sqrt{A (K_u - K_d)}$. For the same value of
  $D_s = 1.74$ pJ/m that optimizes skyrmion stability in antisymmetric
  bilayers, we have $D_c \simeq 3$ mJ/m$^2$ and
  $\Delta \mathcal E \simeq 145 k_BT_{RT}$ at a critical thickness
  $d_c$ finely tuned to around 0.6 nm. The region where the barrier
height remains higher than 50$k_BT_{RT}$ extends over a fairly
broad region of $D_s$ values larger than 0.8 pJ/m and thicknesses
below $d = 1$ nm. The skyrmion solution radii are increasing as
one gets closer to the critical DMI line $D = D_c$.  Note
that due to the absence of a saddle point, the expected skyrmion
behavior close to the critical DMI value $D_c$ will be very
different from that of a skyrmion in an antisymmetric bilayer or
a monolayer close to the bursting line. In particular additional
entropic contributions to the collapse energy barrier
\cite{desplat18,vonmalottki19} would become increasingly important as
$D$ approaches $D_c$ due to the vanishing domain wall stiffness as
$D \to D_c$; such a study would go beyond the scope of the present
paper.

Finally, Figs. \ref{fig:DE}(c) and \ref{fig:DE}(f) show the radii $R$
and effective energy barriers $\Delta \mathcal E$ of skyrmions as a
function of $D_s$ and $d$ for a ferromagnetic monolayer (see figure
caption for details). The region of relative stability
($\Delta \mathcal E \geq 20 k_BT_{RT}$) extends over a much
narrower wedge-shaped zone as compared to the asymmetric bilayer
and SAF. The effective energy barrier reaches up to about
$45k_BT_{RT}$. This, however, requires a very sharp tuning of the
layer thickness $d$ around 0.7 nm and $D_s$ around 2 pJ/m or beyond.

\section{Conclusions}
\label{sec:concl}

Since the first observations of skyrmions and their displacement with
low electric current densities, considerable efforts have been devoted
to the practical realization of order 10 nm radius skyrmions existing
at zero applied magnetic field with the liftime and dynamics
compatible with information technology applications. In this quest,
for a long time the stray field effects have been generally ignored,
considered negligible because of the ultrathin characteristics of the
magnetic layers in which the skyrmions appeared and/or because of the
presence of strong, dominating DMI. In later developments, stray field
effects have been seen as a source of instability, as they promote the
enlargement of skyrmions and their disappearance via bursting or
stripe-out, as observed in many experiments on ferromagnetic mono- and
multilayers.

In the present study, we demonstrated that, to the contrary, the stray
field may exert a pronounced stabilizing effect on magnetic skyrmions.
We established that in exchange-decoupled antisymmetric bilayers the
stray field and the DMI are tailored to be highly favorable to
skyrmion stability through a coordinated action of the DMI and the
dipolar interaction to form a chiral pair of skyrmions in a
flux-closure pattern.

To fully assess the performance of this system, we carried out a
combined analytical and computational micromagnetic study, focusing on
the parameter regime in which the skyrmions have the form of tiny
magnetic bubbles whose shape may be characterized by a single
parameter --- the skyrmion radius. In particular, we obtained an
accurate representation of the stray field energy of a skyrmionic
bubble as a function of its radius that is asymptotically valid for
sufficiently thin films and the radii exceeding the transition layer
thickness of the skyrmionic bubble. In this regime, we obtained zero
magnetic field skyrmion phase diagrams as a function of only two
dimensionless parameters, the reduced DMI strength and the reduced
film thickness, governing the system’s behavior.

The importance of the obtained phase diagrams lies in the fact that
they provide, for the first time, the region of skyrmion stability via
a quantitative prediction of skyrmion bursting, in agreement with the
numerics. They also give a prediction for the effective energy barrier
that determines skyrmion stability against thermal noise and allows to
locate the material parameters for which this stability is optimal.
We wish to point out the very important role of the analytical
modeling in our work for guiding the micromagnetic simulations: it is
the knowlege of the position of the optimal skyrmion stability line
and the use of the skyrmion solution configurations along it as
starting points for the simulations that enabled us to obtain a full
phase diagrams in the non-dimensional parameter space within a
practical computational time.

The non-dimensional phase diagrams can, in turn, be used to predict
the skyrmion size and stability across a broad spectrum of
micromagnetic parameters, as demonstrated in the case of Pt/Co/AlO$_x$
based materials. Using this example, we illustrated the potential of
the exchange-decoupled antisymmetric ferromagnetic layers and compared
it with that of SAF and ferromagnetic monolayers. We found that the
antisymmetric bilayer provides a broadening of the range of existence
of skyrmion solutions, which could potentially result in a significant
decrease of the degree of fine-tuning necessary to obtain very compact
and stable skyrmions.  We hope that our findings, together with the
recent experimental verification of a dipolar-field-enhanced effective
DMI in antisymmetric bilayers, will motivate further experimental
studies into stray-fields-assisted skyrmion stabilization.

\begin{acknowledgments}
  A. Bernand-Mantel was supported by France 2030 government investment
  plan managed by the French National Research Agency under grant
  reference PEPR SPIN ANR-22-EXSP-0009 and grant NanoX
  ANR-17-EURE-0009 in the framework of the Programme des
  Investissements d'Avenir. C. B. Muratov was supported by MUR via
  PRIN 2022 PNRR project P2022WJW9H and acknowledges the MUR
  Excellence Department Project awarded to the Department of
  Mathematics, University of Pisa, CUP I57G22000700001. C. B. Muratov
  is a member of INdAM-GNAMPA. The authors would also like to express
  their gratitude for the hospitality of the Isaac Newton Institute
  within the scope of the program ``Recent challenges in the
  mathematical design of new materials'', during which some of the
  ideas of this work had been conceived.
\end{acknowledgments}

\appendix

\section{Computation of the stray field energy}
\label{sec:appstray}

In this section we compute the non-local surface-surface charge
contribution to the magnetostatic energy for the configuration in
\eqref{eq:bubble}, which is defined by
\begin{align}
  E_{ss}(\mpa) = -{\bar\delta}\int_{\R^2} \int_{\R^2}
  {\nabla \mpa(\mathbf r) \cdot \nabla \mpa(\mathbf r') \over 4
  \pi |\mathbf r - \mathbf r'|}  \, d^2r  \, d^2 r'.
\end{align}
We introduce the 2D Fourier transform 
\begin{align}
  \widehat f(\q) = \int_{\R^2} e^{-i \q \cdot \rr} f(\rr)   \, d^2r, 
\end{align}
and note that
\begin{align}
  E_{ss}(\mpa)
  & = -  \frac{\bar\delta}{2} \int_{\R^2} q \,
    |\widehat{\mpa+1}|^2  \frac{d^2 q}{(2\pi)^2} \notag \\
  & = -  \frac{\bar\delta}{2} \| \mpa + 1 \|^2_{\mathring{H}^{1/2}(\R^2)},
\end{align}
where $q = |\mathbf q|$ and $\| \mpa + 1
\|^2_{\mathring{H}^{1/2}(\R^2)}$ is the homogenous Gagliardo norm
squared of $\mpa + 1$, that may be alternatively defined as
\cite{lieb-loss,dinezza12} 
\begin{align}
  \label{eq:Gagliardomp}
  \| \mpa + 1 \|^2_{\mathring{H}^{1/2}(\R^2)} = \int_{\R^2}
  \int_{\R^2} {(\mpa(\rr) - \mpa(\rr' ))^2 \over 4 \pi |\rr - \rr'|^3}
  d^2 r \, d^2 r'. 
\end{align}

We wish to compute the norm in the expression for $E_{ss}(\mpa)$
asymptotically for $\bar \rho \gg 1$. To this aim, we define
\begin{align}
  \widehat{\mpa+1}
  &= \int_{\R^2} e^{-i {\mathbf q} \cdot {\mathbf r}}
    (\mpa+1) \, d^2 r \notag \\ 
  & = \int_0^{2\pi} \int_0^\infty e^{-i q r \cos\theta} (1+
    \tanh(\bar\rho-r)) r \, dr \, d \theta \notag \\ 
  &= 2\pi  I,
\end{align}
where
\begin{align}
I=\int_0^\infty J_0(q r) [1 + \tanh(\bar\rho - r)] r \, dr,
\end{align}
and here and below $J_n(z)$ is the Bessel function of the first kind
of order $n$ \cite{abramowitz}. Note that $1 + \tanh(\bar\rho - r)$
decays exponentially to $0$ as $r \to \infty$, and hence the above
integral converges absolutely for all $q$ and $\bar\rho$.

We introduce a new variable $y=r/\bar\rho$ and perform an integration
by parts:
\begin{align}
  I
  & =\bar\rho^2 \int_0^\infty y  J_0(q\bar\rho y)  [1 +\tanh(\bar\rho(1 -
    y)) ] \, dy \notag \\
  & = \frac{\bar\rho^2}{q} \int_0^\infty y J_1(q\bar\rho y)\, {\rm
    sech}^2(\bar\rho(1-y)) \, dy,
\end{align}
where we used the identity $[z J_1(z)]' = z J_0(z)$ \cite{abramowitz}.
We then again change variables to $z = \bar\rho(1-y)$ and obtain
$I = \bar \rho I_0 / q$, where
\begin{align}
  \label{eq:I0}
  I_0
  &=\int_{-\infty}^{\bar\rho} \left(1-\frac{z}{\bar\rho}\right)
    J_1(q\bar\rho-qz)\, {\rm sech}^2z \, dz .
\end{align}

For each $q$ fixed and $\bar\rho \gg 1$ the integrand in \eqref{eq:I0}
is exponentially small for all $z > \bar \rho$. Therefore, we can
extend the domain of integration in \eqref{eq:I0} to the whole of $\R$
to obtain that to within $O(e^{-\bar\rho})$
\begin{align} 
  I_0&\simeq\int_{-\infty}^\infty \left(1-\frac{z}{\bar\rho}\right)
       J_1(q\bar\rho-qz)\, {\rm sech}^2 z \, dz. 
\end{align}
Now we utilize the integral representation of the Bessel function
\cite{abramowitz}
\begin{equation}
  \label{eq:J1}
J_1(z) = \frac{1}{\pi} \int_0^\pi \sin \theta \sin(z \sin \theta) \,
d\theta, 
\end{equation}
and change the order of integration, arriving at
\begin{multline}
  I_0 \simeq \frac{1}{\pi} \int_0^\pi \sin \theta \, \bigg\{
  \int_{-\infty}^{\infty} \left(1-\frac{z}{\bar\rho}\right)
  \text{sech}^2 z   \\ \times \sin\left[ (q\bar\rho -
      qz)\sin \theta \right] \, dz \bigg\} \, d \theta.
\end{multline}
Using basic trigonometry, we can write
\begin{align}
  \int_{-\infty}^{\infty}
  &\left(1-\frac{z}{\bar\rho}\right)\text{sech}^2 z \sin[ (q\bar\rho -
    qz)\sin \theta] \, dz =\notag \\ 
  & \sin(q\bar\rho \sin \theta) \int_{-\infty}^{\infty}
    \left(1-\frac{z}{\bar\rho}\right)\text{sech}^2 z \cos(qz \sin
    \theta) \, dz \nonumber \\  
  &- \cos(q\bar\rho \sin \theta) \int_{-\infty}^{\infty}
    \left(1-\frac{z}{\bar\rho}\right)\text{sech}^2 z \sin(qz \sin
    \theta) \, dz. 
\end{align}
Then, due to the even and odd symmetries of the functions involved in
the integrands we can simplify this to
\begin{align}
  \int_{-\infty}^{\infty}
  &\left(1-\frac{z}{\bar\rho}\right)\text{sech}^2 z \sin[ (q\bar\rho -
    qz)\sin \theta] \, dz =\notag \\ 
  & \sin(q\bar\rho \sin \theta) \int_{-\infty}^{\infty}
    \text{sech}^2 z \cos(qz \sin \theta) \, dz \nonumber  \\
  &+\frac{1}{\bar\rho} \cos(q\bar\rho \sin \theta)
    \int_{-\infty}^{\infty} z \,  \text{sech}^2 z \sin(qz \sin
    \theta) \, dz. 
\end{align}

The integrals appearing in the right-hand side of the above expression
are evaluated as
\begin{align}
  & \int_{-\infty}^{\infty} \text{sech}^2 z \cos kz \, dz
    = \frac{\pi
    k}{\sinh(\pi k / 2)}, \\
  & \int_{-\infty}^{\infty}
    z \, \text{sech}^2 z \sin kz \, dz \notag\\
  & \qquad = \frac{\pi}{\sinh(\pi k / 2)} \left[ \frac{\pi k}{2}
    \coth\left(\frac{\pi k}{2}\right) - 1 \right].
\end{align}
Setting $k = q \sin \theta$, we then have $I_0 \simeq I_1+I_2$, where
\begin{align}
  I_1
  & =\int_0^\pi \frac{q \sin^2 \theta \sin(q\bar\rho \sin
    \theta)}{\sinh(\frac{\pi q \sin \theta}{2})} \, d\theta,  \\
  I_2
  & = \frac{1}{\bar\rho}\int_0^\pi
    \frac{ \sin \theta
    \cos(q\bar\rho \sin
    \theta)}{\sinh(\frac{\pi q \sin
    \theta}{2})}   \notag \\ 
  &\qquad \times \left[ \frac{\pi q\sin\theta}{2} \coth\left(\frac{\pi
    q \sin\theta}{2}\right) - 1 \right]\, d\theta. 
\end{align}
Now, for each $q$ fixed and $\bar\rho \gg 1$, we observe that $I_1$
and $I_2$ have highly oscillatory integrands and, therefore, we can
use the stationary phase method (around $\theta=\frac{\pi}{2}$)
\cite[Proposition 3, Chapter 8]{stein} to evaluate the above
integrals. After a few manipulations, we obtain to within
$O(1/\bar\rho)$ relative errors:
\begin{align}
  I_1
  &
    \simeq q \, \mathrm{csch} \left( {\pi q \over 2} \right)
    \int_0^\pi \sin \theta \sin (q \bar \rho \sin \theta) \, d \theta
    \notag \\ 
  & = \pi q\, {\rm csch} \left( \frac{\pi q}{2}  \right)
    J_1(q\bar\rho),
\end{align}
and
\begin{align}
  I_2
  & \simeq  {1 \over \bar \rho} \left[ \frac{\frac{\pi
    q}{2} \coth\left(\frac{\pi q}{2}\right) - 1}{\sinh\left(\frac{\pi
    q}{2}\right)} \right] \int_0^\pi \cos (q \bar \rho \sin \theta)
    \, d \theta \notag \\
  & = \frac{\pi J_0(q\bar\rho)}{\bar\rho} \left[ \frac{\frac{\pi
    q}{2} \coth\left(\frac{\pi q}{2}\right) - 1}{\sinh\left(\frac{\pi
    q}{2}\right)} \right], 
\end{align}
where we used \eqref{eq:J1} and the integral representation
\cite{abramowitz}
\begin{align}
  J_0(z) = {1 \over \pi} \int_0^\pi \cos (z \sin \theta)
  \, d \theta .
\end{align}
Notice that the above approximation for $I_1$ also remains valid for
$q \bar \rho \lesssim 1$, as can be seen by Taylor-expanding the
denominator of the integrand in $q \ll 1$ in this case.

We now observe that for $\bar\rho \gg 1$ the integral $I_2$ is
negligible compared to $I_1$, and hence up to an $O(1/\bar\rho)$
relative error we have
\begin{align}
  I = \frac{\bar\rho I_0 }{q} \simeq \frac{\bar\rho I_1}{q} \simeq \pi
  \bar\rho \, {\rm csch} \left( \frac{\pi q}{2} \right)  J_1(q\bar\rho). 
\end{align}
It follows that
\begin{align}
\widehat{\mpa+1} = 2\pi I \simeq 2 \pi^2 \bar\rho J_1(q \bar\rho) \,
  {\rm csch} \left( \frac{\pi q}{2} \right),
\end{align}
and we have
$\| \mpa +1 \|^2_{\mathring{H}^{1/2}(\R^2)} \simeq 2\pi^3 \bar\rho^2
N$, where
\begin{align}
  N = 
  \int_{0}^\infty q^2
  J_1^2 (q \bar\rho)  {\rm
  csch}^2 \left( \frac{\pi q}{2} \right)
  \, d q. 
\end{align}
We next choose $\sigma>0$ such that $\sigma \ll 1$ and
$\sigma \bar\rho \gg 1$, and split this integral as
\begin{align}
  N&=  \int_{0}^\sigma q^2  J_1^2 (q \bar\rho)  \, {\rm csch}^2 \left( 
     \frac{\pi q}{2}  \right) d q \notag \\ 
   & \qquad+ \int_{\sigma}^\infty q^2  J_1^2 (q \bar\rho) \,  {\rm
     csch}^2 \left( \frac{\pi q}{2}  \right) d q \notag \\ 
   &\simeq \frac{4}{\pi^2} \int_{0}^\sigma   J_1^2 (q \bar\rho)  \, d
     q \notag \\ 
   &\qquad+  \frac{2}{\pi \bar\rho}\int_{\sigma}^\infty q
     \sin^2\left( q\bar\rho-{\pi \over 4} \right)  {\rm csch}^2 \left(
     \frac{\pi q}{2} \right) \, d q 
     \notag \\
   & = N_1 + N_2,
\end{align}
to within an error bounded by a quantity of order
$O(\sigma^3) + O(1/(\sigma \bar\rho^2))$, using the asymptotic formula
\cite{abramowitz}
\begin{align}
  J_1(z) \simeq \sqrt{2 \over \pi z} \sin \left( z - {\pi \over 4}
  \right) , \qquad z \gg 1.
\end{align}
For the second integral, we have for $\bar\rho \gg 1$ 
\begin{align}
  N_2
  & = {1 \over \pi \bar \rho} \int_\sigma^\infty q \left[ 1 - \sin (2
    q \bar \rho)  \right]  \, {\rm csch}^2 \left(
    \frac{\pi q}{2} \right) \, d q \notag \\
  & \simeq  {1 \over \pi \bar \rho} \int_\sigma^\infty q \, {\rm
    csch}^2 \left(  \frac{\pi q}{2} \right) \, d q,
\end{align}
up to an error of order $O(1/(\sigma \bar\rho^2))$.  Thus
\begin{align}
  N
  & \simeq \frac{4}{\pi^2 \bar\rho} \int_{0}^{\sigma \bar\rho}   J_1^2
    (s)  \, d s +  \frac{1}{\pi \bar\rho}\int_{\sigma}^\infty q   \,
    {\rm csch}^2 \left( \frac{\pi q}{2}  \right) d q. 
\end{align}

The second integral above is computable explicitly, and using
$\sigma\ll 1$ we have asymptotically
\begin{align}
  \int_{\sigma}^\infty q   \, {\rm csch}^2  \left( \frac{\pi q}{2}  \right) d q
  & \simeq \frac{4}{\pi^2} \left( 1-\ln\pi - \ln \sigma \right),
\end{align}
to within an $O(\sigma^2)$ error.  The first integral is computable in
terms of the generalized hypergeometric functions, and it is possible
to find its asymptotic behavior to within an $O(1/(\sigma \bar \rho))$
error (arguing, e.g. as in \cite{newman63}):
\begin{align}
  \int_{0}^{\sigma \bar\rho}   J_1^2 (s)  \, d s   \simeq
  \frac{1}{\pi} \left( -2 +\gamma + 3 \ln 2 +\ln \bar\rho + \ln 
  \sigma  \right) . 
\end{align}

\begin{figure}[t]
  \centering
  \includegraphics[width=3.25in]{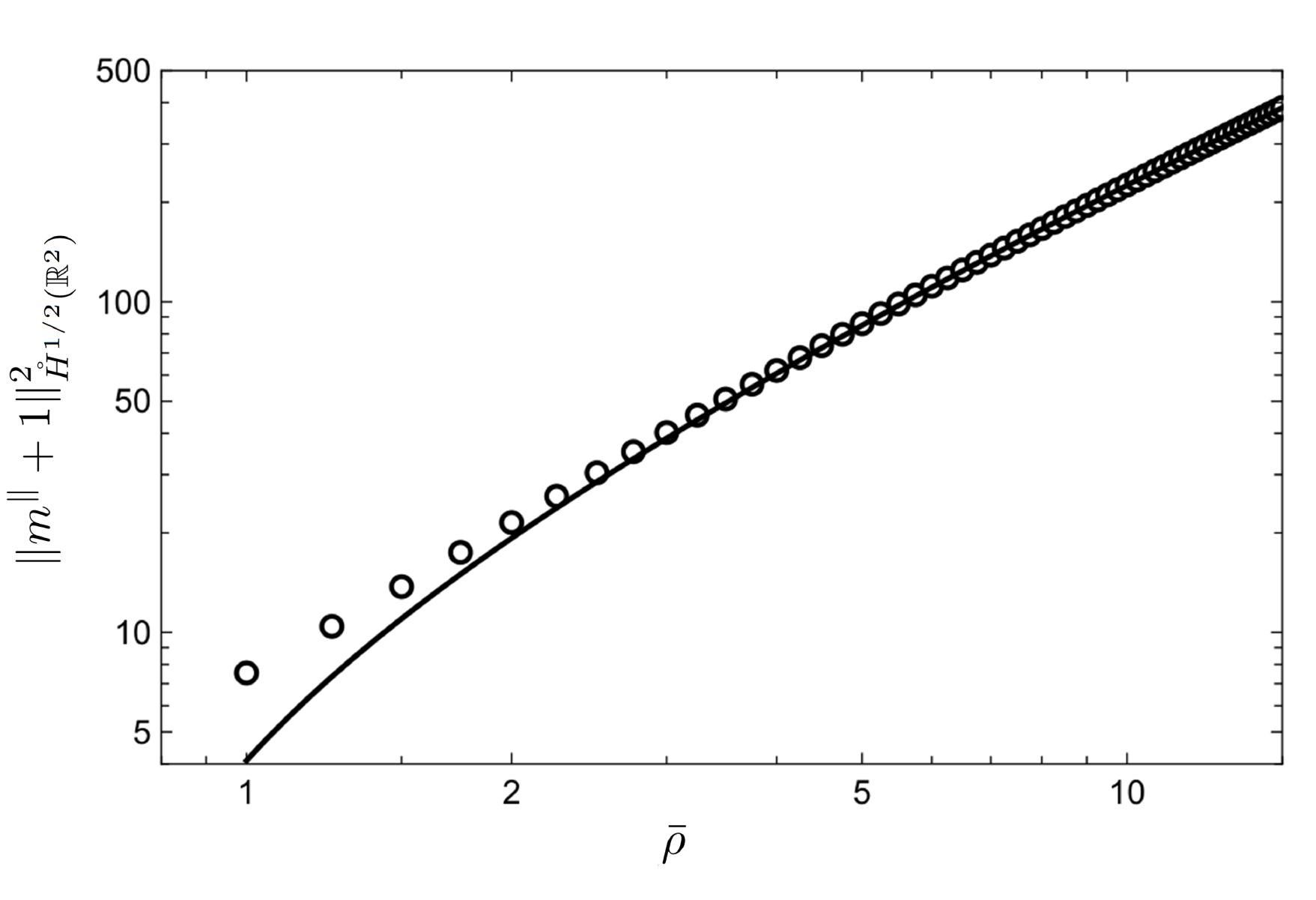}
  \caption{Comparison of the quantity in the left-hand side of
    \eqref{eq:Nlog} (open circles) with the prediction in the
    right-hand side (solid line).}
  \label{fig:converg}
\end{figure}

Using all these results, we finally obtain, up to a relative error of
algebraic order in $1/\bar\rho$:
\begin{align}
  \label{eq:Nlog}
  \| \mpa +1 \|^2_{\mathring{H}^{1/2}(\R^2)}
  &\simeq 8 \bar\rho \ln
    \left(\frac{8\bar\rho}{\pi}
    e^{\gamma-1} \right),
\end{align}
which is one of our principal results. To verify its accuracy, we
evaluated the norm numerically for a range of $\bar\rho$, using the
representation in \eqref{eq:Gagliardomp} and compared the result with
the one given by \eqref{eq:Nlog}. Taking advantage of the radial
symmetry, the computation of the norm may be reduced to the following
double integral amenable to an accurate numerical evaluation
\cite[Eq. (3.617)]{gradshteyn}:
\begin{align}
  \| \mpa
  & +1 \|^2_{\mathring{H}^{1/2}(\R^2)} = \notag \\
  & \int_0^\infty
    \int_r^\infty [\tanh (\bar \rho - r) - \tanh (\bar \rho - r')]^2
    \notag \\
  & \times {4 r r' \, \mathrm E \left(  {2 \sqrt{r r'} \over r' + r}
    \right) \over (r' - r)^2 (r' + r)} \, dr' \, dr,
\end{align}
where
$\mathrm E(k) = \int_0^{\pi/2} \sqrt{1 - k^2 \sin^2 \theta} \, d
\theta$ is the complete elliptic integral of the second kind. A
comparison of the numerical results and the prediction of the formula
in \eqref{eq:Nlog} is presented in Fig. \ref{fig:converg}. One can see
that already for $\bar \rho \geq 2$ the formula in the right-hand side
of \eqref{eq:Nlog} predicts the value of the norm to within
$\sim 10$\% accuracy, and the accuracy rapidly increases with
$\bar\rho$.

\bibliography{bilayer}

\end{document}